\documentclass[a4paper,11pt]{article}
\pdfoutput=1

\usepackage{jheppub}

\usepackage{amsmath,latexsym,amssymb,slashed}
\usepackage{bbm}
\usepackage{multirow}

\newcommand\beal{\begin{align}}
\newcommand\nn{\nonumber}

\newcommand{\eq}[1]{\begin{equation}#1\end{equation}}
\newcommand{\spl}[1]{\begin{split}#1\end{split}}
\newcommand{\al}[1]{\begin{align}#1\end{align}}
\newcommand{\all}[1]{\begin{align*}#1\end{align*}}

\newcommand{\arx}[1]{\href{http://arxiv.org/abs/#1}{[{\tt #1}]}}

\newcommand{\arxth}[1]{\href{http://arxiv.org/abs/hep-th/#1}{[{\tt hep-th/#1}]}}

\newcommand{\ccal}{\mathcal{C}}
\newcommand{\hcal}{\mathcal{H}}
\newcommand{\mcal}{\mathcal{M}}
\newcommand{\ncal}{\mathcal{N}}
\newcommand{\fcal}{\mathcal{F}}
\newcommand{\scal}{\mathcal{S}}
\newcommand{\bs}{\left.\right|_{\Sigma}}

\newcommand{\G}{\Gamma}
\newcommand{\g}{\gamma}
\newcommand{\e}{\epsilon}

\newcommand{\p}{\partial}
\renewcommand{\o}{\omega}
\renewcommand{\O}{\Omega}
\newcommand{\m}{\mu}
\newcommand{\n}{\nu}
\renewcommand{\k}{\kappa}
\renewcommand{\l}{\lambda}
\renewcommand{\a}{\alpha}
\renewcommand{\b}{\beta}
\newcommand{\s}{\sigma}
\renewcommand{\t}{\theta}
\renewcommand{\L}{\Lambda}
\newcommand{\z}{\zeta}
\newcommand{\gs}{\left.g\right|_{\Sigma}}

\def\d{\text{d}}

\def\slashchar#1{\setbox0=\hbox{$#1$}           
\dimen0=\wd0                                 
\setbox1=\hbox{/} \dimen1=\wd1               
\ifdim\dimen0>\dimen1                        
\rlap{\hbox to \dimen0{\hfil/\hfil}}      
#1                                        
\else                                        
\rlap{\hbox to \dimen1{\hfil$#1$\hfil}}   
/                                         
\fi}

\title{Supersymmetric solutions on $SU(4)$-structure deformed Stenzel space}

\author{D.L.A. Prins}

\affiliation{
Universit\'{e} de Lyon\\
UMR 5822, CNRS/IN2P3, Institut de Physique Nucl\'{e}aire de Lyon\\
4 rue Enrico Fermi, F-69622 Villeurbanne Cedex, France\\
}
\emailAdd{dlaprins@ipnl.in2p3.fr}

\abstract{
The Stenzel space fourfold is a non-compact Calabi-Yau which is a higher dimensional analogue of the deformed conifold. We consider $\ncal = (1,1)$ type IIA, $\ncal = 1$ M-theory and $\ncal= (2,0)$ type IIB compactifications on this Stenzel space, thus examining the gravity side of potentially higher dimensional analogues of Klebanov-Strassler-like compactifications. We construct families of $SU(4)$-structures and solve associated moduli spaces, of complex and symplectic structures amongst others. By making use of these, we can construct IIA compactifications on manifolds homeomorphic to the Stenzel space fourfold, but with complex non-CY $SU(4)$-structures. Such compactifications are sourced by a distribution of NS5-branes. The external metric is asymptotically conformal $\text{AdS}_3$ and should thus be suitable for holography applications.
}

\begin{document}
\maketitle
\flushbottom
\setcounter{footnote}{0}
\renewcommand{\thefootnote}{\arabic{footnote}}
\setcounter{section}{0}

\section{Outline}
When compactifying type II supergravity or M-theory, one usually splits up the ten or eleven dimensional space into a direct product of an `external' and an `internal' manifold. The former is interpreted as spacetime and is typically required to be either AdS or Minkowski space. As fluxless supersymmetric vacua in 4+6 dimensions require the internal space to be Calabi-Yau, these have become a benchmark of sorts. The internal space can be taken to be either compact or non-compact; advantages of the latter are that explicit Calabi-Yau metrics are known only for non-compact spaces, and that non-compact supersymmetric flux vacua circumvent the no-go theorem of \cite{malnun}.

A class of such non-compact Calabi-Yau manifolds are Stenzel spaces \cite{stenzel}: particular cases are Eguchi-Hanson space \cite{eguchi} for $d=4$ and the deformed conifold \cite{candelas} for $d=6$. The conifold $\ccal(T^{1,1})$ has been of particular interest, as the (warped) metric of a compactification on $\mathbbm{R}^{1,3} \times \ccal(T^{1,1})$ asymptotes to $\text{AdS}_5 \times T^{1,1}$, thus lending itself to describe dual conformal field theories \cite{kw}, \cite{kt}. Although the conifold has a singularity at the tip of the cone, this may be smoothed out by either deformation (blow-up) or resolution. Smoothness at the tip of the `throat' of the deformed conifold is then associated to color confinement in the dual CFT \cite{ks}.

In \cite{cglp}, an $\ncal=2$ M-theory flux vacuum was constructed on $\mathbbm{R}^{1,2} \times \scal$, where $\scal$ is the fourfold in the family (hereafter simply referred to as `Stenzel space'). This M-theory vacuum was reduced to an $\ncal = (2,2)$ vacuum of type IIA theory in \cite{kp}. Similarly to the deformed conifold, Stenzel space is a deformed cone, in this case over the Stiefel manifold $V_{5,2} \simeq SO(5)/SO(3)$, with an $S^4$ bolt at the origin to smooth out the singularity of $\ccal(V_{5,2})$. Hence this vacuum is a higher dimensional (or lower, depending on perspective) analogue of the one found in \cite{ks}. Recently, this vacuum has attracted interest mostly in the context of metastable vacua in M-theory when supplemented with (anti-) $M_2$ branes \cite{kp, bena, massai}. See also \cite{hashimoto, ms}.

In order to construct holographic field theory duals, it is not necessary that the internal metric is Calabi-Yau. On the gravity side, more general vacua can be found that are not necessarily Calabi-Yau. A useful tool in the construction of such spaces are $G$-structures \cite{gauntlett}. $G$-structures describe geometrical structures on manifolds, and classify obstructions to integrability of these algebraically in terms of intrinsic torsion. In particular, a manifold is Calabi-Yau\footnote{Our definition for a Calabi-Yau $n$-fold here is that the holonomy is a (not necessarily proper) subgroup of $SU(n)$, rather than the stricter definition that the holonomy is exactly $SU(n)$.} if and only if it has an $SU(n)$-structure with vanishing intrinsic torsion.
Recently, we studied vacua on spaces with $SU(4)$-structures for type IIB \cite{ptb} as well as type IIA and M-theory \cite{pta} and found classes of $\ncal = (2,0)$ IIB, $\ncal = (1,1)$ IIA vacua. The latter uplift to $\ncal = 1$ M-theory vacua\footnote{We abuse terminology somewhat here; these vacua are merely solutions to the equations of motions of $D=11$ supergravity which admit supersymmetric branes.}.

The purpose of this paper is three-fold. First of all, we are interested in extending the results of \cite{cglp} to construct IIA, IIB and M-theory vacua with less supersymmetry, by applying the results of \cite{ptb, pta} in the case of vanishing torsion. These vacua allow for more RR fluxes, which can be described in terms of certain closed and co-closed $(p,q)$-forms. Thus, this process involves constructing closed and co-closed forms on Stenzel space and considering their effects on the warp factor. For the case of IIA, we find three new possible contributions: scalar terms, a primitive $(1,1)$-form, and a new primitive $(2,2)$-form. Although massive IIA contributes such scalar terms, non-massive IIA may also do so. All of these terms cause divergences in the warp factor, the scalars in the UV, the (1,1) and (2,2)-form in the IR. Thus, in this sense, these vacua are more along the lines of \cite{kw} than that of \cite{ks}. We explain how such fluxes affect the uplift of the IIA vacua to M-theory vacua on $\mathbbm{R}^{1,2} \times \scal$.

Secondly, we are interested in constructing explicit examples of non-Calabi-Yau spaces. Stenzel space comes equipped with a natural $SU(4)$-structure with vanishing intrinsic torsion defined by its Calabi-Yau structure (i.e., its symplectic form, holomorphic four-form, and metric). We use the coset structure of $V_{5,2}$ to construct a family of $SU(4)$-structures which we call `left-invariant', as these are induced from the left-invariant forms on $SO(5)/SO(3)$. We consider a sub-family of these, which we refer to as `$abc$ $SU(4)$-structures', and give explicit formulae for the torsion classes. This reduces the problem of finding moduli spaces, e.g. of integrable almost complex structures, to ODE which we solve. We will abuse terminology somewhat and refer to manifolds diffeomorphic to Stenzel space equipped with a different $SU(4)$-structure as `$SU(4)$-deformed Stenzel spaces'.

Our third point of interest is to construct vacua on such $SU(4)$-deformed Stenzel space. Due to the specifics of forms on Stenzel space, non-Calabi-Yau deformations automatically violate the NSNS Bianchi identity. We construct type IIA $\ncal= (1,1)$ vacua on complex non-symplectic (hence, in particular, non-Calabi-Yau) $SU(4)$-deformed Stenzel space, up to subtleties in the integrability theorem. The geometry is smooth and complete, with an $S^4$ bolt at the origin and conical asymptotics, thus conformal to $\text{AdS}_3$ in the UV; this is similar to IIA on Stenzel space. The RR flux is primitive, (2,2) and satisfies the RR Bianchi identity. The violation of the NSNS Bianchi is sourced by a distribution of NS5-branes, for which we give the contribution to the action. These vacua do not uplift to M-theory vacua on $\mathbb{R}^{1,2} \times \scal$, as such M-theory vacua require certain constraints on the dilaton in terms of the warp factor which are explicitly not satisfied.

The rest of this paper is organized as follows. In section \ref{prelim}, we discuss some known results on Stenzel space, $SU(4)$-structures, and vacua on manifolds with $SU(4)$-structures. In section \ref{fluxstenzel}, we describe the combination of the ingredients of the previous section, applying the $SU(4)$-solutions to Stenzel space. We also describe calibrated probe D-branes on the IIA Stenzel space vacua. In section \ref{deforms} we describe how to construct $SU(4)$-structures on manifolds diffeomorphic to Stenzel space.  We then examine the moduli spaces and geodesical completeness of such spaces. Finally, in section \ref{generalizations} we discuss how to construct IIA vacua on these $SU(4)$-deformed Stenzel spaces. This involves taking the susy solutions of section \ref{prelim}, applying them with torsion classes described in section \ref{deforms}, checking integrability conditions and Bianchi identities. In appendix A we discuss results when considering type IIB on both Stenzel space and $SU(4)$-deformed Stenzel space. In appendix B, some interesting geometrical features are highlighted which can arise when one allows for violations of the RR Bianchi identities in addition to the NSNS Bianchi identities.

\section{Preliminaries}\label{prelim}
\subsection{Stenzel Space}
The vacua of interest will either have a Stenzel space or something closely related to a Stenzel space as internal manifold. A Stenzel space can be viewed as a smoothing of a (singular) cone in such a way that it is still a Calabi-Yau (CY) manifold. Stenzel spaces exist for arbitrary even dimensions; the most well-known is the deformed conifold in $d=6$ \cite{candelas}. We will focus purely on the case where $d=8$ and will simply refer to the CY fourfold Stenzel space as `Stenzel space'. We will review its properties below; see \cite{stenzel}, \cite{cglp}, \cite{kp} for more details.

Stenzel space is defined as the CY manifold $(\scal, J, \O)$, where $\scal$ is a smooth manifold, $J$ its K\"{a}hler form, and $\O$ the Calabi-Yau form.
There are numerous ways to describe Stenzel space. As an algebraic variety, it is defined by the set
\al{
\scal = \{ z \in \mathbbm{C}^5 \; | \; z^2 = \epsilon^2 \}
}
with $\epsilon \in \mathbbm{R}$.  Topologically speaking, it is homeomorphic to $T^*S^4$. Alternatively, and for our purposes more conveniently, Stenzel space can be considered as a deformed cone over the Stiefel manifold $V_{5,2} \simeq SO(5) / SO(3)$, with the singularity at the tip of the undeformed cone blown up to $S^4 \simeq SO(5) / SO(4)$. It is thus a smooth non-compact manifold. $SO(5)$ comes equipped with a set of left-invariant forms $L_{AB}$, $A,B \in \{1, ...,5\}$ satisfying the relation
\al{
\d L_{AB} = L_{AB} \wedge L_{BC} \;.
}
Relabeling $A =(1,2, j)$, $j \in \{1,2,3\}$ and defining $L_{1j} = \s_j$, $L_{2j} = \tilde{\s}_j$, $L_{12} = \n$, one has that $\n, \s_j, \tilde{\s}_j$ span a basis for $T^*V_{5,2}$ and satisfy the following relations (summation implied):
\eq{\spl{\label{eq:liforms}
\d \s_j &= \n \wedge \tilde{\s}_j + L_{jk} \wedge \s_k \\
\d \tilde{\s}_j &= - \n \wedge \s_j + L_{jk} \wedge \tilde{\s}_k \\
\d \n &= - \s_j \wedge \tilde{\s}_j \;.
}}
Stenzel space comes equipped with a metric
\al{\label{eq:stenzelmetric}
ds^2(\scal) = c(\tau)^2 \left( \frac14 d \tau^2 + \n^2 \right) + b(\tau)^2 \tilde{\s}_j^2 + a(\tau)^2 \s_j^2
}
with $a,b,c$ defined as
\eq{\spl{\label{eq:stenzelabc}
a^2 &= 3^{-\frac14} \l^2 \e^{\frac32}  x \cosh \left(\frac{\tau}{2}\right) \\
b^2 &= 3^{-\frac14} \l^2 \e^{\frac32}  x \cosh \left(\frac{\tau}{2}\right) \tanh^2 \left(\frac{\tau}{2}\right) \\
c^2 &= 3^{\frac34} \l^2 \e^{\frac32}  x^{-3} \cosh^3 \left(\frac{\tau}{2}\right)\;,
}}
with
\al{
x \equiv (2 + \cosh \tau)^{1/4}
}
defined for convenience. $\tau \in [0, \infty)$ is the radial coordinate of the deformed cone. Such $a,b,c$ satisfy the differential constraints
\eq{\spl{\label{eq:cycondition}
a'&= \frac{1}{4 b} \left(b^2 + c^2 - a^2 \right)  \\
b'&= \frac{1}{4 a} \left(a^2 + c^2 - b^2 \right) \\
c'&=\frac{3 c}{4 a b} \left(a^2 + b^2 - c^2 \right)
}}
which were determined to be the constraints for Ricci-flatness in \cite{cglp}. Note that $\l \in \mathbbm{R}$ is an arbitrary scaling parameter, as Ricci-flatness is invariant under conformal transformations.
We define orthonormal one-forms
\eq{\begin{alignedat}{2}\label{eq:oncoords}
e_0 =& \frac{c(\tau)}{2} \d\tau \;, \quad &e_j = a(\tau) \s_j \\
\tilde{e}_0 =& c(\tau) \n \;, \quad &\tilde{e}_j= b(\tau) \tilde{\s}_j
\end{alignedat}}
and holomorphic one-forms
\eq{\spl{ \label{eq:holocoords}
\z^0 &= - \frac{c}{2}d\tau +i c \n \\
\z^j &= a \s_j + i b \tilde{\s}_j \;.
}}
Using these, the K\"{a}hler form and Calabi-Yau form are given by
\eq{\spl{ \label{eq:stenzelsu}
J &= \frac{i}{2} \zeta^\a \wedge \bar{\z}^\a  \\
\O &= \z^0 \wedge\z^1 \wedge \z^2 \wedge \z^3
}}
such that $(J, \O)$ forms a Calabi-Yau structure on $\scal$. For small $\tau$, one sees that $a \sim c \sim \l \e^{3/4}$, $b \sim \frac{1}{2} \l \e^{3/4} \tau$. Hence at $\tau = 0$,  the CY-structure degenerates to
\eq{\spl{
\left.ds^2(\scal)\right|_{\tau = 0} &= \l^2 \e^{3/2} \left( \n^2 + \s_j^2 \right) \\
\left. J \right|_{\tau = 0} &= 0 \\
\left. \O\right|_{\tau = 0} &= \l^4 \e^3 i \n \wedge \s_1 \wedge \s_2 \wedge \s_3 \;.
}}
In particular, the metric becomes the standard metric on $S^4$, $J$ vanishes as the $S^4$ bolt is a special Lagrangian submaniold, and $\O$ becomes proportional to the volume form of the bolt.

\subsection{Type II vacua on spaces with strict $SU(4)$-structures}
In this section a summary is given of previous results on supersymmetric vacua of type II supergravity and M-theory which possess a strict $SU(4)$-structure. For more details, see \cite{ptb,pta}.

\subsubsection{$SU(4)$-structures}\label{SU4}
Let $\mcal_8$ be an oriented eight-dimensional Riemannian manifold. An $SU(4)$-structure can be defined as a real two-form $J$ and a complex four-form $\O$, satisfying
\eq{\spl{
J \wedge \O &= 0 \\
\frac{1}{4!} J^4 = \frac{1}{2^4} \O \wedge \O^* &= \text{vol}_8 \;.
}}
These define a compatible almost complex and almost symplectic structure, as well as a trivialization of $\bigwedge^{(4,0)} T^{*+}\mcal_8$, with $T^{*+} \mcal_8$ the $+i$-eigenbundle of the almost complex structure induced on the cotangent bundle. Geometrical structures and obstructions to their existence are defined through torsion classes $W_1,...,W_5$. More specifically,
\eq{\spl{
\d J =& W_1 \lrcorner \O^* + W_3 + W_4 \wedge J + \text{c.c.} \\
\d \O =& \frac{8 i}{3} W_1 \wedge J \wedge J + W_2 \wedge J + W_5^* \wedge \O \;.
}}
Here, $W_1, W_4, W_5$ are complex (1,0)-forms, while $W_2, W_3$ are complex primitive\footnote{A three-form $W$ of $\mcal_8$ is primitive if and only if $W \wedge J \wedge J  = 0$. By Hodge duality, this is equivalent to $W_{mnp} J^{np} = 0$. Occasionally, this is referred to as `tracelesness'. We will most certainly not do so. } (2,1)-forms. As can be seen, the almost symplectic structure is a symplectic structure if and only if $W_1 = W_3 = W_4$. Slightly less straightforwardly, one can confirm that the almost complex structure is integrable if and only if $W_1 = W_2 = 0$ by checking whether or not the Nijenhuis tensor vanishes.
We have summarized the for us relevant geometrical structures in the table below.
\begin{center}
\begin{tabular}{|l|l|}
\hline
Geometrical structure & Torsion classes \\ \hline
Complex & $W_1 = W_2 = 0$ \\ \hline
Symplectic & $W_1 = W_3 = W_4 = 0$ \\ \hline
K\"{a}hler & $W_1 = W_2 = W_3 = W_4 = 0$ \\\hline
Nearly Calabi-Yau & $W_1 = W_3 = W_4 = W_5 = 0$ \\ \hline
Conformal Calabi-Yau & $W_1 = W_2 = W_3 =0$, $2 W_4 =  W_5$ \\\hline
Calabi-Yau & $W_j = 0 $ $\forall j$ \\\hline
\end{tabular}
\end{center}

\subsubsection{Type II theory on manifolds with $SU(4)$-structures}\label{typeII}
We now turn our attention to the physical side of things. We consider warped vacua of the type
\al{
ds^2(\mcal_{10}) = e^{2A} ds^2(\mathbbm{R}^{1,1}) + ds^2 (\mcal_8)\;.
}
We will refer to $\mathbbm{R}^{1,1}$ as the external space and $\mcal_8$ as the internal.
We will also make use of the notation
\all{
e^{2A} \equiv \hcal^{-1}
}
 where this is more convenient.
We make use of a democratic formulation of type II supergravity (see \cite{mmlt} for our conventions), where the RR fluxes in type IIA/B are given by
\eq{\spl{ \label{eq:selfduality}
\fcal &= \sum_{p \in I_{A/B}} \fcal_{(p)} \qquad \left.
\begin{array}{cl}
I_A &= \{0,2,4,6,8,10\}\\
I_B &= \{1,3,5,7,9\}
\end{array} \right. \\
\fcal &=  \star \s \fcal \;,
}}
with $\fcal_{(p)}$ a $p$-form, the map $\s$ reversing the order of indices (i.e., $\s = (-1)^{\frac12 p (p-1)}$), and the twisted selfduality constraint imposed at the level of the equations of motion. Two-dimensional Lorentz invariance of the vacuum, combined with the twisted selfduality constraint,  means that the RR fluxes can be further decomposed as
\al{
\fcal &= e^{2A} \text{vol}_2 \wedge \star_8 \s F + F \;,
}
where $F$ will be referred to as the `magnetic' RR flux.

Existence of an $SU(4)$-structure on $\mcal_8$ implies that all tensors, a priori irreducible representations of $SO(8)$, can be globally decomposed into irreducible $SU(4)$ representations. From a more geometrical point of view, it is equivalent to think of this procedure in terms of a Hodge decomposition and a decomposition into primitive forms (i.e., decompositions with respect to the almost complex and almost symplectic structure)\footnote{Note also that this is another way to think of intrinsic torsion: $\d J, \d\O$ are tensors which can be decomposed into $SU(4)$-irreps due to the $SU(4)$-sructure.}. We do so for the fluxes, leading to the following decomposition:
\eq{\spl{\label{eq:decomp}
F_m =& f^{(1,0)}_{1|m} + f^{(0,1)}_{1|m} \\
F_{mn}=& f^{(1,1)}_{2|mn}+f_2J_{mn}+\left(f^{(2,0)}_{2|mn}+\mathrm{c.c.}\right)\\
F_{mnp} =& F_{mnp}=f^{(2,1)}_{3|mnp}+3f^{(1,0)}_{3|[m}J_{np]} +\tilde{f}^{(1,0)}_{3|s}\Omega^{s*}{}_{mnp} +\mathrm{c.c.}\\
F^+_{mnpq} =& f^{(2,2)}_{4|mnpq}+6f_{4}J_{[mn}J_{pq]} +\left(6f^{(2,0)}_{4|[mn}J_{pq]}+\tilde{f}_4\Omega_{mnps} + \mathrm{c.c.}\right)\\
F^-_{mnpq} =& 6f^{(1,1)}_{4|[mn}J_{pq]}+\left(f^{(3,1)}_{4|mnpq}+\mathrm{c.c.}\right)\;.
}}
Here, $F^{\pm}$ refer to the selfdual and antiselfdual parts of the four-form. $p$-forms with $p>0$ can be dualized and then decomposed similarly. The NSNS three-form  can first be split into an external and internal part,
\al{\label{eq:hdecomp}
H = e^{2A}\text{vol}_2 \wedge H_{1} + H_3\;,
}
after which the the one- and three-form $H_1$, $H_3$ will be decomposed similarly.

A supersymmetric vacuum is a vacuum that is invariant under some supersymmetry. Supersymmetric vacua are found by solving certain Killing spinor equations, called the supersymmetry equations. In our conventions, these are given by
\eq{\spl{\label{kse}
\delta \lambda^1 &= \left(\underline{\p}\phi + \frac{1}{2} \underline{H} \right) \epsilon_1+ \left( \frac{1}{16} e^{\phi} \G^M \underline{\fcal}\G_M \G_{11} \right) \epsilon_2 = 0 \\
\delta \lambda^2 &= \left( \underline{\p }\phi - \frac{1}{2} \underline{H} \right) \epsilon_2 - \left( \frac{1}{16} e^{\phi} \G^M \sigma(\underline{\fcal})\G_M \G_{11} \right) \epsilon_1 = 0 \\
\delta \psi^1_M &= \left( \nabla_M + \frac{1}{4} \underline{H}_M \right) \epsilon_1 + \left( \frac{1}{16} e^{\phi} \underline{\fcal}\G_M \G_{11} \right) \epsilon_2 = 0 \\
\delta \psi^2_M &= \left( \nabla_M - \frac{1}{4} \underline{H}_M \right) \epsilon_2 - \left( \frac{1}{16} e^{\phi} \sigma(\underline{\fcal})\G_M \G_{11} \right) \epsilon_1 = 0 \;.
}}
In order for solutions to these susy equations to satisfy the equations of motion (referred to as `integrability' of the susy equations), the integrability theorem described in \cite{pta} shows that there are two additional requirements: the Bianchi identities
\al{
\d_H \fcal = 0 \;, \quad \d H =0
}
must be satisfied, and the equation of motion obtained by varying the action with respect to $B$, formally $\delta H_{MN} =0$, must be explicitly satisfied for $MN = 01$.

In order to solve the susy equations, one can decompose the Killing spinor $\epsilon$ into an internal and an external part. Here the $SU(4)$-structure comes into play once more: existence of an $SU(4)$-structure on $\mcal_8$ is equivalent to the existence of a pure spinor, which can most easily be seen due to the so-called accidental isomorphism $SU(4) \simeq Spin(6)$. By making use of this pure spinor in the ansatz for the susy Killing spinor paramater (a so called `strict $SU(4)$' ansatz), one can solve the susy equations. For type IIB with $\ncal = (2,0)$, this Killing spinor ansatz is given by
\eq{\spl{
\e_1 &= \a \left(\z \otimes \eta + \z^c \otimes \eta^c \right) \\
\e_2 &= \a \left(\z \otimes e^{i \t} \eta + \z^c \otimes e^{- i \t} \eta^c \right)
}}
and the solution to the supersymmetry equations is then given by
\eq{\spl{\label{eq:iib}
W_1 &= W_2 = 0 \\
W_3 &= i e^\phi (\cos\t f^{(2,1)}_3  - i \sin\t f^{(2,1)}_5) \\
W_{4} &= \frac{2}{3}\p  (\phi- A )\\
W_{5} &=  \p (\phi - 2 A +  i \t)\\
\a &= e^{ \frac{1}{2} A}\\
\tilde{f}^{(1,0)}_{3} &=  \tilde{f}^{(1,0)}_{5} = \tilde{h}^{(1,0)}_{3} =0 \\
h_{1}^{(1,0)} &= 0\\
h_{3}^{(1,0)} &= \frac{2}{3} \p \t \\
f_{1}^{(1,0)} &=-i \p ( e^{- \phi}\sin\t )\\
f_{3}^{(1,0)} &= - \frac{i}{3}e^{2A} \p ( e^{-2A- \phi}\cos\t )\\
f_{5}^{(1,0)} &=\frac{1}{3}e^{-4A} \p( e^{4A- \phi}\sin\t )\\
f_{7}^{(1,0)} &=e^{-2A} \p( e^{2A- \phi}\cos\t )\\
h^{(2,1)} &= e^\phi (- \cos\t f^{(2,1)}_5  + i \sin\t f^{(2,1)}_3)
~.
}}
Here, $\phi$ is the dilaton and $\a$, $\t$ are parameters in the Killing spinor ansatz.
Note that the almost complex structure is integrable ($W_1= W_2 = 0$), hence the Dolbeault operator $\p$ is well-defined.

For type IIA with $\ncal = (1,1)$, the strict $SU(4)$ Killing spinor ansatz is given by
\eq{\spl{\label{eq:iiaks}
\e_1 &= \frac{\a}{\sqrt{2}} \z_+ \otimes\left( \eta + \eta^c \right) \\
\e_2 &= \frac{\a}{\sqrt{2}} \z_- \otimes \left( e^{i \t} \eta +  e^{- i \t} \eta^c \right) \;.
}}
Depending on the phase $\t$, three branches of solutions are found. For each branch, one finds that
\eq{\spl{\label{eq:hint}
\alpha &= e^{ \frac12 A}\\
H_1 &= - 2\d A~,}}
with the warp factor $A$ a function on $\mcal_8$, while the magnetic fluxes obey a (twisted) self-duality condition:
\eq{\spl{\label{eq:11fluxes1}
F=\star_8\sigma F
~.}}
In addition, the RR fluxes obey the following relations:
\eq{\spl{  \label{eq:rrconstraints}
f_4&=\frac16 f_0+\frac43 e^{-i\theta}\cos\theta\tilde{f}_4\\
f_2&=2e^{-i\theta}\sin\theta\tilde{f}_4 \\
\sin \t f_{2|mn}^{(2,0)} &= - \cos \t f_{4|mn}^{(2,0)} - \frac{1}{8} e^{i \t} \O_{mn}^{\phantom{mn}pq} f_{4|pq}^{(0,2)} \\
\sin \t f_{4|mn}^{(2,0)} &=   \cos \t f_{2|mn}^{(2,0)} - \frac{1}{8} e^{i \t} \O_{mn}^{\phantom{mn}pq} f_{2|pq}^{(0,2)}
\;,}}
where $f_0$, $f_2$, $f_4$ are real scalars while $\tilde{f}_4$ is complex. Note that the last two equations are equivalent for $e^{2i \t} \neq 1$, whereas for $e^{2 i \t} = 1$ they become independent pseudoreality conditions. For the remaining NS fields, we distinguish between three cases.\\

$\bullet$ $e^{2i\t} = 1$:
\eq{
\spl{\label{eq:iiasol1}
e^\phi &=  g_s e^A\\
h_{3}^{(1,0)} = \tilde{h}_{3}^{(1,0)} &= 0\\
h_{3}^{(2,1)} &=0\\
W_1 &=-\frac{3i}{4}W_4\\
W_3 &=\frac{1}{2}W_2\\
W_5 &=\frac{3}{2}W_4
~,}}
with $g_s$ a non-zero integration constant, $A$, $W_4^{(1,0)}$, $W_2^{(2,1)}$ unconstrained.\\

$\bullet$ $e^{2i\t} = -1$:
\eq{
\spl{\label{eq:iiasol2}
h_{3}^{(1,0)} &= 0\\
\tilde{h}_{3}^{(1,0)} &=  \frac{1}{4}\partial^{+}(A-\phi)\\
W_1 &=0\\
W_{2} &=-2i h_{3}^{(2,1)}\\
W_3 &=0\\
W_4 &=\partial^{+}(\phi - A)\\
W_5 &=\frac{3}{2}\partial^{+}(\phi - A)
~,}}
with $A$, $\phi$,  $h_3^{(2,1)}$ unconstrained; for any scalar $S$, $\partial^{\pm}S$ denotes the projection of the exterior derivative $\d S$ onto its (1,0), (0,1) parts.\\

$\bullet$ $e^{2i\t} \neq \pm 1$:
\eq{
\spl{\label{eq:iiasol3}
e^\phi&=g_s e^{A}\cos\theta \\
h_{3}^{(1,0)} &= \frac{2}{3}\partial^{+}\t\\
\tilde{h}_{3}^{(1,0)} &= \frac{1}{4}(i+\tan\t)\partial^{+}\t\\
W_1^{(1,0)}&=\frac{1}{4}(1+i\cot\t)\partial^{+}\t\\
W_{2}^{(2,1)} &=2(-i+\cot\t)~\!h_{3}^{(2,1)}\\
W_3^{(2,1)}&=\cot\t ~\!h_{3}^{(2,1)}\\
W_4^{(1,0)}&=-(\tan\t+\frac13\cot\t)\partial^{+}\t\\
W_5^{(1,0)}&=(i-\frac{1}{2}\cot\t-\frac{3}{2}\tan\t)\partial^{+}\t
~,}}
with $g_s$ a non-zero integration constant,  $A$, $\t$,  $h_3^{(2,1)}$ unconstrained.

\subsubsection{Type IIA $\ncal = (1,1)$ Calabi-Yau vacua}\label{IIACY}
In order to find explicit CY vacua, we restrict the solution to $W_j = 0$ $\forall j$ and explicitly apply the Bianchi identities and the integrability constraint. The constraints then found are as follows.
The metric is given by
\al{\label{eq:iiametric}
ds^2(\mcal_{10}) = \hcal^{-1} ds^2(\mathbbm{R}^{1,1}) + ds^2(\mcal_8)\;.
}
The NSNS three-form, dilaton and warp factor are constrained to be
\eq{\spl{\label{nsf}
e^{\phi} &= g_se^A \\
H &=-\mathrm{vol}_2\wedge\d e^{2A}\;,
}}
while the magnetic RR forms are given by
\eq{\spl{\label{rrf}
F_0&=f_0\\
F_2&=f_2 J + f_2^{(1,1)}+ \left(f_2^{(2,0)} + \text{c.c.}\right)\\
F_4&= \star_8F_4 = f_4^{(2,2)} + f_4 J \wedge J + \left( \tilde{f}_4 \O + f^{(2,0)}_4 \wedge J + \text{c.c} \right)
\\
F_6&=-\star_8F_2\\
F_8&=\star_8F_0
~,
}}
where $J$ and $\O$ are the K\"{a}hler form and holomorphic four-form of the Calabi-Yau fourfold respectively. In addition, the RR fluxes obey the constraint \eqref{eq:rrconstraints} with $\t$ now constant.

This supersymmetry solution automatically satisfies the NSNS Bianchi identity, i.e., $\d H=0$. The RR Bianchi identity reduces to closure and co-closure of the magnetic RR flux:
\eq{\label{eq:cybianchi}
\d F=\d\star_8 F=0~.}
More explicitly this means that $f_0$, $f_2$, $f_4$, $\tilde{f}_4$ are constant while $f_2^{(1,1)}$, $f_2^{(2,0)}$, $f_4^{(2,0)}$, $f_4^{(2,2)}$ are closed and co-closed. The integrability constraint reduces to:
\eq{\label{int2}
-\d\star_8\d e^{-2A}+\frac{g_s^2}{2} F\wedge \sigma(F)|_8=0
~.
}
which is more conveniently written as
\al{\label{eq:warp}
\nabla^2 \hcal = -  g_s^2\star_8 \left(  F_0 \wedge \star_8 F_0 + F_2 \wedge \star_8 F_2 + \frac{1}{2}F_4 \wedge \star_8 F_4 \right)\;,
}
the Hodge dual being taken with respect to $ds^2(\mcal_8)$ as this equation is purely internal.

\subsubsection{M-theory $\ncal = 1$ Calabi-Yau vacua}\label{m-theory}
Let us examine uplifts of the type IIA $\ncal= (1,1)$ vacua to M-theory vacua on $\mathbbm{R}^{1,2} \times \mcal_8$. In order to lift the type IIA vacua, we set $F_0 = F_2=0$; the former because massive type IIA does not lift, the latter because we are not interested in fibrations. We also require $e^{2 i \t} = 1$ in order for the Killing spinor to properly lift. Note that this restricts the torsion classes of such $d=3$, $\ncal = 1$ vacua to satisfy
\eq{\spl{\label{eq:mtorsion}
W_1 = - \frac{3 i}{4} W_4 &=- \frac{i}{2} W_5 \\
W_3 = \frac{1}{2} W_2 \;.
}}
For reasons that will become clear later on, we are only interested in the case with $W_1 = W_3 = 0$, thus immediately restricting ourselves to the Calabi-Yau case. Such CY solutions are as follows.
\\
\\
The warped metric is given by
\eq{\label{eq:11metric}
ds^2=e^{\frac{4}{3}A}ds^2(\mathbb{R}^{1,2})+e^{- \frac{2}{3} A} ds^2(\mcal_8)
~,}
where now $\mcal_8$ is a CY fourfold.
The four-flux
is given by
\eq{\spl{
G &=\mathrm{vol}_3\wedge\d(e^{2A})+F  \\
F &= f_4 \left(J \wedge J +\frac{3}{2}~\!\mathrm{Re}\O \right)+f_4^{(2,2)} +  \left(f^{(2,0)}_4 + \text{c.c}\right) \wedge J \;.
}}
The Bianchi identity and equation of motion for $G$ (or, equivalently, the uplift of the IIA Bianchi identities) require that $F$ is closed and co-closed:
\eq{\label{44}\d F=\d\star_8 F=0~,}
hence $f_4$ is a (real) constant, while $f_4^{(2,2)}$ is harmonic.
Similarly, the warp factor is constrained to satisfy
\al{
-\d\star_8\d e^{-2 A} +\frac12 F\wedge F=0~.
}
Again, this is more conveniently rewritten as
\al{ \label{eq:mwarp}
\nabla^2 \hcal = - \star_8 \left( \frac{1}{2}F \wedge \star_8 F\right) .
}

\section{Type IIA \& M-theory on Stenzel Space}\label{fluxstenzel}

\subsection{$\ncal=(1,1)$ IIA Fluxes on Stenzel Space}\label{sec:iia}
We will now consider the case of IIA theory on Stenzel space. Stenzel space possesses an $SU(4)$-structure as it is CY, hence applying the constraint given in subsection \ref{IIACY} yields a solution to the susy equations on Stenzel space. The only obstructions to finding a supersymmetric vacuum come from the fact that the (magnetic) RR fluxes have to be closed and co-closed on the $\scal$ and that the warp factor has to satisfy an inhomogeneous Laplace equation whose source is given by the RR flux. So the strategy in the following will be to a) construct new closed and co-closed RR fluxes on Stenzel space and b) solve the resulting differential equation for the warp factor.

In \cite{cglp}, an $\ncal = 2$ M-theory vacuum was given on Stenzel space. The vacuum was defined by a primitive selfdual closed (2,2)-form
\al{\label{eq:stenzelL2}
f^{(2,2)} &=      3 f_L [\tilde{e}_0\wedge e_1 \wedge e_2 \wedge e_3 + e_0 \wedge \tilde{e}_1 \wedge \tilde{e}_2 \wedge \tilde{e}_3]\nn\\
&+ \frac{1}{2} f_L \e_{ijk}[e_0\wedge e_i \wedge e_j \wedge \tilde{e}_k +  \tilde{e}_0 \wedge e_k \wedge \tilde{e}_i \wedge \tilde{e}_j ]\;,
}
with
\al{
f_L \equiv \left(\e^3 \cosh^4 \frac{\tau}{2}\right)^{-1}
}
and the associated warp factor
\al{\label{eq:stenzeln=2}
\hcal = 2^{3/2} 3^{11/4} \e^{-9/2} \int^\infty_x \frac{dt}{ (t^4 -1)^{5/2} }
}
which satisfies  \eqref{eq:mwarp}. Clearly, such an $\ncal=2$ M-theory vacuum is also an $\ncal=1$ M-theory vacuum as described in section
\ref{m-theory}, and can be dimensionally reduced to an $\ncal = (1,1) $ IIA  vacuum, as it satisfies \eqref{eq:warp} with $F_0 = F_2 = 0$, $F_4 = f^{(2,2)}$.

More generally, for $\ncal = (1,1) $ IIA  vacuum there are  four different irreps to consider: scalars, primitive $(1,1)$-forms, $(2,0)$-forms and primitive $(2,2)$-forms.
An observation that simplifies the analysis is that the contribution of each irrep to the warp factor equation \eqref{eq:warp} can be considered separately. For example,
switching on a scalar flux $f_s$ leads to the following modification of the warp factor equation
\all{
\nabla^2 \hcal &= -\frac{1}{2} \star_8 \left(f^{(2,2)}\wedge  f^{(2,2)} \right) - f_s^2\;.
}
So without loss of generality we can simply set $\hcal= \hcal_{(2,2)} + \hcal_s$, provided that
\all{
\nabla^2 \hcal_s =  -f_s^2
~;~~~\nabla^2 \hcal_{(2,2)} = -\frac{1}{2}\star_8 \left(f^{(2,2)}\wedge  f^{(2,2)} \right)
~.
}
The upshot is that each irreducible component of the flux $f$ can be considered as a separate source to the warp factor equation, leading to a corresponding solution $\hcal_f$ for the warp factor; the complete expresion for the warp factor is then given by $$\hcal = \sum_f \hcal_f~.$$

{\bf Summary of the results:}
We have been able to construct a non-normalizable closed and co-closed primitive $(1,1)$-form and a non-normalizable closed and co-closed primitive $(2,2)$-form on Stenzel space, and for each of them we have calculated the contribution to the warp factor. We find divergences in the IR\footnote{A note on terminology: `IR' for us means $\tau \rightarrow 0$, whereas `UV' is synonymous with $\tau \rightarrow \infty$. For a function $X$, we will also occasionally use `$X_{IR}$', which is defined by $X \rightarrow X_{IR}$ when $\tau \rightarrow 0$,  and similarly for $X_{UV}$ and $\tau \rightarrow \infty$.} which cannot be canceled against each other; in the UV the warp factor goes to zero in exactly the same way as for \eqref{eq:stenzeln=2}.

In addition we have considered closed and co-closed scalar fluxes $f_0$, $f_2$, $f_4$, $\tilde{f}_4$, which appear through the Romans mass ($f_0$), but also through the two form ($F_2\sim f_2 J+\dots$) and the four-form ($F_4\sim f_4 J^2+(\tilde{f}_4\Omega+\mathrm{c.c.})+\dots$). These scalars are constant as follows from closure and co-closure, and hence the corresponding fluxes are  non-normalizable. We have calculated the induced contribution to the warp factor: in the IR the warp factor stays finite; in the UV the geometry becomes  singular at a  finite value of the radial coordinate.

Finally we also consider adding a homogeneous solution $\hcal_0$ to the warp factor:   $\nabla^2\hcal_0=0$. This is finite in the UV but diverges in the IR and for that reason it was discarded in \cite{cglp}. Here we are including divergent modifications to the warp factor, so it is consistent to take  $\hcal_0$  into account.

\subsubsection{Scalars}
All scalars will contribute in the same manner but with different prefactors. Specifically, we have
\al{
  f^2 &\equiv  \star_8 \left(  F_0 \wedge \star_8 F_0 + F_2 \wedge \star_8 F_2 + \frac{1}{2}F_4 \wedge \star_8 F_4 \right)\nn\\
&=   \left(f_0^2 +4 f_2^2 + 12 f_4^2 + 16|\tilde{f}_4|^2\right)\;,
}
which gives the source entering the warp factor equation (\ref{eq:warp}).
Using the definitions of the metric \eqref{eq:stenzelmetric} and $a,b,c$ \eqref{eq:stenzelabc}, it can be seen that the warp equation reduces to
\all{
\hcal' &= - \varphi \left(x\sinh\left(\frac{\tau}{2}\right) \right)^{-3}\int^\tau dt \sinh^3 \!t\\
 &=  - \varphi \left(x\sinh\left(\frac{\tau}{2}\right) \right)^{-3}\left(\frac{2}{3}-\cosh\tau +\frac{1}{3}\cosh^3 \!\tau\right)\\
 &=  - \varphi \left(x\sinh\left(\frac{\tau}{2}\right) \right)^{-3}\left(\frac{4}{3}-4\cosh^4\left(\frac{\tau}{2}\right) +\frac{8}{3}\cosh^6 \left(\frac{\tau}{2}\right)\right)
~,}
where we fixed the integration constant in the second line by demanding that there should not be any IR divergence and in the third line we used the identity
$\cosh \tau = 2 \cosh^2\left(\frac{\tau}{2}\right) -1$; we have also defined
\all{
\hcal' &= \p_\tau \hcal \\
\varphi &\equiv g_s^2 f^2 (\frac{3^{3/4} }{2} \l^2 \epsilon^{3/2}) > 0\;.
}
By noting that
\all{
\left(\frac{4}{3}-4\cosh^4\left(\frac{\tau}{2}\right) +\frac{8}{3}\cosh^6\left(\frac{\tau}{2}\right)\right) = \frac{4}{3}x^{4}\sinh^4\left(\frac{\tau}{2}\right)
~,}
we find the explicit expression for the warp factor:
\al{
\hcal = - \frac{8\sqrt{2}}{15}\varphi \int_{3^{5/4}}^{x^5}\frac{dt}{(t^{4/5}-1)^{1/2}}+k\;,
}
with $k$ an integration constant.
There are no singularities in the IR since $\tau \rightarrow 0$ corresponds to  $x^5 \rightarrow 3^{5/4}$. In the UV, on the other hand, $\hcal$ vanishes at a finite
value of the radial coordinate and the metric develops a singularity. More specifically the asymptotics are:
\eq{
\hcal\longrightarrow
\left\{\begin{array}{rl}
k~;&~~~\tau \rightarrow 0\\
\frac{3}{4} k(\tau_{\mathrm{UV}}-\tau)~;&~~~\tau \rightarrow \tau_{\mathrm{UV}}
\end{array}
\right.
~,}
where
\eq{
\tau_{\mathrm{UV}}\equiv\frac{4}{3}\log \big(
\frac{2^{\frac14}9k}{8\varphi}
\big)~,
}
and we have assumed $\tau_{\mathrm{UV}} >>1$.

The ten- and eleven-dimensional metrics\footnote{The solution can be uplifted to eleven dimensions provided the Romans mass vanishes.} are given in \eqref{eq:iiametric} and \eqref{eq:11metric} respectively. In the IR the metric is regular and asymptotes to the standard metric on $\mathbb{R}^{1,1}\times \scal$, or $\mathbb{R}^{1,2}\times \scal$ after uplift to eleven dimensions. In the UV the metric becomes singular at $\tau=\tau_{\mathrm{UV}}$; the singularity remains even after the uplift.

The dilaton and NSNS flux are given in (\ref{nsf}). The RR flux is given in (\ref{rrf}); in the present case of scalar fluxes this gives:
\eq{\spl{\label{rrscalar}
\mathcal{F}_0&=f_0\\
\mathcal{F}_2&=\hcal^{-1}f_0\mathrm{vol}_2+f_2J\\
\mathcal{F}_4&=\hcal^{-1}f_2J\wedge\mathrm{vol}_2+f_4J\wedge J+(\tilde{f}_4\Omega+\mathrm{c.c.})
~,
}}
where $f_0$, $f_2$, $f_4$ are constants.

\subsubsection{Two-forms}\label{sec:twoforms}
We now consider the possibility of allowing two-forms. There are two cases: $f_2^{(2,0)}$ or $f_2^{(1,1)}$, since  $f_4^{(1,1)}$ vanishes and $f^{(2,0)}_4$ is determined by $f^{(2,0)}_2$. For simplicity we will henceforth set $f^{(2,0)}=0$. The Bianchi identities require that $F_2$ should be closed and co-closed. Since $\star F_2 \sim f^{(1,1)}_2 \wedge J \wedge J$  and since $J$ is closed, closure of $F_2$ also implies co-closure.

In terms of the holomorphic one-forms \eqref{eq:holocoords}, a closed and co-closed primitive $(1,1)$-form is given by
\eq{\label{eq:(1,1)1}
f^{(1,1)} = i f(\tau) ( \z^0\wedge \bar{\z}^{\bar{0}} - \frac{1}{3} \z^j\wedge \bar{\z}^{\bar{\jmath}})\;,
}
provided that
\al{\label{eq:(1,1)2}
2 a b f' + (3 c^2 +2 (ab)') f = 0\;.
}
Using the explicit expressions for $a,b,c$ \eqref{eq:stenzelabc}, we find that
\eq{\label{ft}
f(\tau) = \frac{m}{\left(x\sinh\left(\tau/2\right)\right)^{4}}\;,
}
with $m$ an integration constant.
The norm of $f^{(1,1)}_2$ can be calculated explicitly:
\all{
F_2 \wedge \star F_2 &= -\frac{1}{2}f^{(1,1)}_2 \wedge f^{(1,1)}_2 \wedge J \wedge J \\
&= \frac{1}{3}f^2 \O \wedge \O^* = \frac{16}{3} f^2 \text{vol}_8 \;,
}
with $f$ given in (\ref{ft}). Note that this flux is non-normalizable: one has that
\eq{\spl{
\int_{\mcal_8} F_2 \wedge \star F_2 &= \int d^8 x \sqrt{g} \frac{16}{3} f^2\\
&\sim  \lim_{\e \rightarrow 0} \int_\e^\infty d\tau \frac{\sinh^3 \tau}{x^8 \sinh^8\left(\tau/2\right)}
~,}}
hence the integral diverges in the IR as $\tau^{-4}$.

The warp factor is given by
\all{
\nabla^2 \hcal = - g_s^2\frac{16}{3}f^2\;,
}
which leads to
\eq{\label{hder1}
 \hcal =  \frac{2}{3^{5/4}} g_s^2 \l^2 \e^{3/2} \int_{\tau}^{\infty}\d t\; \frac{k  \left( x \sinh\left(\frac{t}{2}\right)\right)^4 - m^2 }{\left( x \sinh\left(\frac{t}{2}\right)\right)^{7}}\;,
}
where $k$ is an integration constant, $m$ was introduced in (\ref{ft}), and the limits of the integration have been chosen so that $\hcal$ vanishes in the UV, but diverges in the IR.

Explicitly, in the UV one has:
\eq{
\hcal(\tau)\longrightarrow  \frac{128}{6^{1/4}27}g_s^2\l^2 \e^{3/2}  k e^{-\frac{9}{4}\tau}~;~~\tau \rightarrow \infty
~.}
In the IR, one finds that
\al{
\hcal(\tau)\longrightarrow  \frac{4}{1215} g_s^2 \l^2 \e^{3/2} \left( -480 m^2 \tau^{-6} + 420 m^2 \tau^{-4} + \frac12(497 m^2 - 540 k) \tau^{-2} \right)
}
for $\tau \rightarrow  \tau_{IR}$. $\tau_{IR}$ is determined by the approximate vanishing of $\hcal$, which happens for
\all{
\tau_{IR} =  2 \sqrt{ - \frac{105 m^2 \pm m\sqrt{32400 k - 18795 m^2 }} {540 k - 497 m^2}}\;.
}
In order for this approximation to be sensible, this places bounds on $k$ as a function of $m$. Specifically, let us take
$\sqrt{\frac{m^2}{k} } << 1$. Then, one has that
\al{
\tau_{IR} = \frac{2}{\sqrt{3}} \left(\frac{m^2}{k}\right)^{1/4} \left( 1 - \frac{7}{24} \sqrt{ \frac{m^2}{k}} \right) + \mathcal{O} ( (\frac{m^2}{k})^{5/4} )\;,
}
which goes to 0 as it should to justify the approximation used for $\hcal$.
In the IR the ten-dimensional string-frame metric becomes singular.
In the UV the string metric asymptotes to conformal\footnote{
Alternatively, the metric can be rewritten as
\all{
ds^2 = \L^2 \left( \d R^2 + R^6 ds^2(\mathbb{R}^{1,1}) + R^2 ds^2(V_{5,2})\right)\;.
}
For this reason, conformal AdS is equivalent to a domain wall.
} AdS.
\eq{
ds^2= \L^2 e^{2 \rho} \left[ds^2(\mathrm{AdS}_3)+ds^2(V_{5,2}) \right]
~,}
with
\eq{\spl{\label{eq:confads}
ds^2(\mathrm{AdS}_3) &= \d \rho^2 + e^{4 \rho}  ds^2(\mathbb{R}^{1,1}) \\
ds^2(V_{5,2}) &= \frac{9}{16} \n^2 + \frac{3}{8} \left( \s_j^2 + \tilde{\s}_j^2 \right)\;.
}}
Here, we introduced
\eq{\label{eq:confadspara}
\rho \equiv \frac{3}{8} \tau \;, \qquad \L^2 \equiv \frac{8}{6^{5/4}} \l^2 \epsilon^{3/2}
}
and rescaled
\all{
ds^2(\mathbb{R}^{1,1}) \rightarrow \frac{32}{9} \L^{4} g_s^{2} k\; ds^2(\mathbb{R}^{1,1}) \;.
}
The Einstein metric $ds^2_E=e^{-\frac{\phi}{2}}ds^2$ is also conformal to $\mathrm{AdS}_3\times V_{5,2}$ but with a different conformal factor:
\eq{
ds^2_E= \left(\frac{32 k}{9}\right)^{1/4}  \L^{5/2} e^{\frac{1}{2} \rho} \left[ds^2(\mathrm{AdS}_3)+ds^2(V_{5,2})\right]
~.}
After uplift to eleven dimensions the metric asymptotes $\mathrm{AdS}_{4}\times V_{5,2}$, exactly as was the case in \cite{cglp}.

The dilaton and NSNS flux are given in (\ref{nsf}). The RR flux is given in (\ref{rrf}); in the present case of (1,1) fluxes this gives:
\eq{\spl{\label{rr11}
\mathcal{F}_0&=0\\
\mathcal{F}_2&=f^{(1,1)}\\
\mathcal{F}_4&=\hcal^{-1}f^{(1,1)}\wedge\mathrm{vol}_2
~,
}}
where $f^{(1,1)}$ was given in (\ref{eq:(1,1)1}), (\ref{ft}).

\subsubsection{Four-forms}
Let us now examine possible four-forms on Stenzel space. The $(3,1)$, $(4,0)$ and non-primitive $(2,2)$ forms have been discussed, so let us examine possible $(2,2)$ forms which are primitive, closed (and hence, co-closed due to selfduality). We have found two such forms; they are given by
\al{
f^{(2,2)}_{NL}& =
            3 f_{NL} [e_0\wedge e_1 \wedge e_2 \wedge e_3 + \tilde{e}_1 \wedge \tilde{e}_2 \wedge \tilde{e}_3 \wedge \tilde{e}_0]\nn\\
&+ \frac{1}{2} f_{NL}  \e_{ijk}[e_0\wedge e_i \wedge \tilde{e}_j \wedge \tilde{e}_k + e_j \wedge e_k \wedge \tilde{e}_i \wedge \tilde{e}_0]
\label{eq:(2,2)NL1}\\
f_L^{(2,2)} &=
     3 f_L [\tilde{e}_0\wedge e_1 \wedge e_2 \wedge e_3 + e_0 \wedge \tilde{e}_1 \wedge \tilde{e}_2 \wedge \tilde{e}_3]\nn\\
&+ \frac{1}{2} f_L \e_{ijk}[e_0\wedge e_i \wedge e_j \wedge \tilde{e}_k +  \tilde{e}_0 \wedge e_k \wedge \tilde{e}_i \wedge \tilde{e}_j ]\;, \label{eq:(2,2)L1}
}
with $f_{NL}$ determined by
\eq{\spl{\label{eq:(2,2)NL2}
2 b c f_{NL}' + ( ac + 6 cb' + 2 bc') f_L =& 0\\
2 c f_{NL} \left(a^2 - b^2 + 2 b a'- 2 a b' \right) =& 0
}}
and $f_L$ determined by
\eq{\spl{\label{eq:(2,2)L2}
2 a c f_{L}' + ( bc + 6 ca' + 2 ac') f_L =& 0\\
2 c f_{L} \left(a^2 - b^2 + 2 b a'- 2 a b' \right) =& 0
}}
Notice that \eqref{eq:(2,2)L2} is nothing more than \eqref{eq:(2,2)NL2} under the transformation $a \leftrightarrow b$. The second equation is a consistency condition which is satisfied for the $a,b,c$ of Stenzel space\footnote{In fact, it is equivalent to integrability of the almost complex structure. See section \ref{moduli} for more details.}. Plugging in $a, b, c$ in the first equation, we find that $f^{(2,2)}_L$ is nothing more than the $L^2$ harmonic form found before \eqref{eq:stenzelL2}. Inserting the expressions for $a,b,c$ \eqref{eq:stenzelabc}, the solution to \eqref{eq:(2,2)NL2} is given by
\al{\label{4}
f_{NL} = \frac{m}{\e^3\sinh^4(\tau/2)}\;,
}
where we have chosen a normalization to match the conventions of \cite{kp}.
This flux is non-normalizable as can be seen from
\all{
\int_{\mcal_8} f^{(2,2)} \wedge f^{(2,2)} &\sim  \lim_{\e \rightarrow 0} \int_\e^\infty d\tau \frac{\sinh^3 \tau}{\sinh^8(\tau/2)}\;,
}
which diverges as $\tau\rightarrow 0$ in the same way as $f^{(1,1)}$ of section \ref{sec:twoforms}. Thus we get the same IR divergence for the warp factor to leading order, although the subleading term will have a different coefficient.
More explicitly the warp factor satisfies
\eq{\spl{
\nabla^2 \hcal &= - \frac{g_s^2}{2} 24 f^2_{NL} \\
&=  - 12 g_s^2\frac{ m^2}{\e^6\sinh^8(\tau/2)}\;,
}}
with solution
\al{
\hcal &= \frac{3^{7/4} }{2}g_s^2 \l^2 \e^{-9/2}\int_{\tau}^{\infty}\d t\;  \frac{(k -2 m^2 \sinh^{-2}\left(\frac{t}{2}\right) - m^2 \sinh^{-4}\left(\frac{t}{2}\right))}
{(x \sinh\left(\frac{t}{2}\right))^{3}} \;,
}
with $k$ an integration constant, and we have chosen the limits of the integration so that $H$ vanishes in the UV but diverges in the IR. Explicitly the UV asymptotics are given by
\al{
\hcal(\tau)\longrightarrow \frac{32}{6^{1/4}} g_s^2 \l^2 \e^{-9/2}  k e^{-\frac{9}{4} \tau}~;~~~\tau \rightarrow \infty
}
and the IR asymptotics by
\eq{
\hcal(\tau)\longrightarrow g_s^2 \l^2\e^{-6}\left( - 32 m^2\tau^{-6}  - 4 m^2 \tau^{-4} + (6k + \frac{223}{30} m^2) \tau^{-2}   \right)
}
for $\tau \rightarrow \tau_{IR}$.
Setting this expression to zero we find
\all{
\tau_{IR} = 2 \sqrt{ \frac{15 m^2 + \sqrt{15 m^2}\sqrt{720 k + 907 m^2}}{180 k + 223 m^2}}\;.
}
Again, we take $k$ such that $\sqrt{\frac{m^2}{k}} << 1$ leading to
\al{
\tau_{IR} = \frac{2}{3^{1/4}} \left(\frac{m^2}{k}\right)^{1/4} \left( 1 + \frac{1}{8 \sqrt{3}}\sqrt{\frac{m^2}{k}}\right)+ \mathcal{O} ( (\frac{m^2}{k})^{5/4} )\;.
}
Similarly to the case where the flux was given by a (1,1)-form, the string metric asymptotes to conformal AdS in the UV
\eq{
ds^2= \L^2 e^{2 \rho} \left[ds^2(\mathrm{AdS}_3)+ds^2(V_{5,2}) \right]
~,}
with $\L$, $\rho$ defined in \eqref{eq:confadspara}, the metrics defined in \eqref{eq:confads}, and the metric rescaled
\all{
ds^2(\mathbb{R}^{1,1}) \rightarrow 24 \L^{4} g_s^{2}\epsilon^{-6} k \;ds^2(\mathbb{R}^{1,1}) \;.
}
Thus, we again find that the Einstein metric is conformal $\mathrm{AdS}_3\times V_{5,2}$ but with a different conformal factor
\eq{
ds^2_E=(24 k)^{1/4}\e^{-3/2}  \L^{5/2} e^{\frac{1}{2}\rho}\left[ds^2(\mathrm{AdS}_3)+ds^2(V_{5,2})\right]
~,}
and again, the uplift to eleven dimensions of the metric asymptotes to $\mathrm{AdS}_{4}\times V_{5,2}$.

The dilaton and NSNS flux are given in (\ref{nsf}). The RR flux is given in (\ref{rrf}); in the present case of (2,2)-fluxes this gives:
\eq{\spl{\label{rr22}
\mathcal{F}_0&=0\\
\mathcal{F}_2&=0\\
\mathcal{F}_4&=f^{(2,2)}
~,
}}
where $f^{(2,2)}$ is given by \eqref{eq:(2,2)NL1},  (\ref{4}).

\subsubsection{Homogeneous solution}\label{sec:homo}
As already  mentioned, the homogeneous solution was discarded in \cite{cglp} due to the IR divergences but for us it is consistent to include it. The warp factor equation becomes
\all{
\p_\tau [\left(x\sinh\left(\frac{\tau}{2}\right)\right)^{3}\p_\tau \hcal] =0 \;,
}
which gives
\al{\label{eq:homo}
\hcal =   k_1 -k_2 \int^\tau \frac{d t}{\left(x(t)\sinh\left(\frac{t}{2}\right)\right)^{3}} \;,
}
with $k_1$, $k_2$ integration constants.
This is finite in the UV but diverges in the IR. Explicitly
the asymptotics are given by
\eq{ \def\arraystretch{1.8}
\hcal \longrightarrow
\left\{\begin{array}{cc}
k_1 +\frac{4}{3^{3/4}} k_2 \tau^{-2} & \phantom{aaaaaa}\tau \rightarrow 0\\
k_1 + \frac{64}{2^{1/4} 9}k_2 e^{-\frac{9}{4}\tau}& \phantom{aaaaaaa}\tau \rightarrow \infty
\end{array}
\right.
~.}

\subsection{Calibrated probe branes on Stenzel Space}\label{caliDbranes}

The purpose of this section is to determine which are the admissible calibrated probe branes on the vacuum $\ncal=(1,1)$ IIA geometry discussed in the previous section.

Consider a supersymmetric vacuum of type II supergravity characterized by two ten-dimensional Majorana-Weyl spinors $\epsilon_1$, $\epsilon_2$ so that $\epsilon_1$ is of positive chirality and  $\epsilon_2$ is of negative, positive chirality in IIA, IIB respectively. We define the polyform
\eq{
\Psi\equiv-\sum_{k=0}^{10}\frac{1}{k!}~\!(\tilde{\epsilon}_1\Gamma_{M_1\dots M_k}\epsilon_2)~\!\d x^{M_1}\wedge\dots\wedge\d x^{M_k}
~,}
the vector
\eq{
K^M\equiv-\frac12 \left( \tilde{\epsilon}_1\Gamma_{M}\epsilon_1+  \tilde{\epsilon}_2\Gamma_{M}\epsilon_2\right)
~,}
and the one-form
\eq{
\omega_M\equiv-\frac12 \left( \tilde{\epsilon}_1\Gamma_{M}\epsilon_1-  \tilde{\epsilon}_2\Gamma_{M}\epsilon_2\right)
~.}
Moreover consider a probe Dp-brane with $(p+1)$-dimensional worldvolume $\Sigma$ and worldvolume electromagnetic fieldstrength\footnote{Not to be confused with the total RR fieldstrength $\fcal$. In the rest of this subsection, we will only discuss the worldvolume fieldstrength, whereas in the rest of the text, we will only discuss the RR fieldstrength. } $\fcal_{\text{wv}}$ whose Bianchi identity reads:
\eq{\label{wvbianchi}
\d\fcal_{\text{wv}}=\left.H\right|_{\Sigma}
~.}
The necessary and sufficient conditions for the $Dp$-brane to be calibrated can then be formulated as follows \cite{melec}:
\eq{\spl{\label{nscal}
\left(\d x^M\wedge\Psi\wedge e^{\fcal_{\text{wv}}}\right)\left.\right|_{\Sigma}&=K^M\sqrt{-\det(\gs+ \fcal_{\text{wv}})}~\!\d^{p+1}\!\sigma\\
\left(\iota_M\Psi\wedge e^{\fcal_{\text{wv}}}\right)\left.\right|_{\Sigma}&=\omega_M\sqrt{-\det(\gs+ \fcal_{\text{wv}})}~\!\d^{p+1}\!\sigma
~,}}
where $\left.g\right|_{\Sigma}$ is the induced metric on the worldvolume and $\sigma^{a}$, $a=1,\dots,p+1$, are coordinates of  $\Sigma$; in  both  left-hand sides above it is understood that we only keep the top $(p+1)$-form.

Let us note two important corollaries that follow from (\ref{nscal}). Firstly it can be shown that the electric worldvolume field defined by $\mathcal{E}\equiv\iota_K \fcal_{\text{wv}}$ is constrained to satisfy:
\eq{\label{cor1}
\mathcal{E}=\omega\bs
~.}
Secondly it can be shown that the vector $K$ belongs to the tangent space of $\Sigma$:
\eq{\label{cor2}
K\in T\Sigma
~.}

\subsection*{Worldvolume equations of motion}

Let the worldvolume be parameterized by $\sigma^{\alpha}$, $\alpha=1,\dots,p+1$.
The dynamical fields on the worldvolume of the branes are the embedding coordinates $X^M(\sigma)$ and the gauge field $A_{\alpha}$. Varying the D$p$-brane action,
\eq{\label{dp}
S_{Dp}= -\mu_p\int_{\Sigma} e^{-\phi}
\sqrt{-\det(\gs+\fcal_{\text{wv}})}~\!\d^{p+1}\!\sigma
+\mu_p\int_{\Sigma}C\wedge e^{\fcal_{\text{wv}}}
~,
}
with respect to $X^M(\sigma)$ we obtain
\eq{\label{xeom}
\partial_{\beta}P_M^{\beta}=0~,}
where we have defined
\eq{\spl{\label{defp}
P_M^{\beta}&\equiv e^{-\phi}\sqrt{-G} \left(G^{(\alpha\beta)}g_{M\alpha} + G^{[\alpha\beta]}B_{M\alpha} \right)
-\frac{1}{p!}\varepsilon^{\beta\alpha_1\dots\alpha_p}\left[\iota_M(C\wedge e^B)\wedge e^{\fcal_{\text{wv}}-B} \right]_{\alpha_1\dots\alpha_p}\\
G_{\alpha\beta}&\equiv \left(g\bs\right)_{\alpha\beta} + \left(\fcal_{\text{wv}}\right)_{\alpha\beta}
~;~~~g_{M\alpha}\equiv g_{MN}\partial_{\alpha}X^N~;~~~
B_{M\alpha}\equiv B_{MN}\partial_{\alpha}X^N
~.}}
Varying (\ref{dp}) with respect to $A_{\alpha}$ we obtain
\eq{\label{aeom}
\partial_{\beta}\Pi^{\alpha\beta}=0~,}
where we have defined
\eq{\label{defpi}
\Pi^{\alpha\beta}\equiv e^{-\phi}\sqrt{-G}
~\!G^{[\alpha\beta]}-\frac{1}{p!}~\!\varepsilon^{\alpha\beta\gamma_1\dots\gamma_p}\left(
C\wedge e^{\fcal_{\text{wv}}}\right)_{\gamma_1\dots\gamma_p}
~.}

\subsubsection*{$\ncal=(1,1)$ IIA CY vacua}
For the  $\ncal=(1,1)$ IIA CY vacua of section \ref{typeII} the two ten-dimensional Majorana-Weyl Killing spinors are given by \eqref{eq:iiaks},
where $\eta$ is a unimodular covariantly-constant spinor of $\mcal_8$  and $\alpha^2=e^A=\hcal^{-1/2}$. It is then straightforward to compute:
\eq{\label{data}
\Psi = \hcal^{-1/2}\left( 1+\hcal^{-1}\d t\wedge\d x \right)\wedge \varphi~;~~~
K =\frac{\partial}{\partial t} ~;~~~
\omega=-\hcal^{-1}\d x
~,
}
where we have defined
\eq{
\varphi\equiv\mathrm{Re}\left[e^{i\theta}(\Omega+e^{iJ})\right]
~.}
Provided a $Dp$-brane is calibrated, i.e. (\ref{nscal}) is satisfied,
its action takes the form:
\eq{\label{dpaction}
S_{Dp}=-\mu_p\int_{\Sigma}\Big(\frac{1}{g_s}~\!\d t\wedge\varphi-C\Big)\wedge e^{\fcal_{\text{wv}}}
~,}
where $\mu_p=(2\pi)^{2p}(\alpha')^{-\frac{p+1}{2}}$ and we have used the fact that  the DBI part of the action saturates the BPS bound
\eq{\label{dbi}
S^{\mathrm{DBI}}_{Dp}= -\mu_p\int_{\Sigma} e^{-\phi}
\sqrt{-\det(\gs+\fcal_{\text{wv}})}~\!\d^{p+1}\!\sigma
=-\mu_p\int_{\Sigma} \frac{1}{g_s}~\!\d t\wedge\varphi \wedge e^{\fcal_{\text{wv}}}
~,}
with $e^{\phi}=g_s \hcal^{-1/2}$.

From (\ref{cor2}) and (\ref{data}) it follows that a calibrated $D$-brane
must extend along the time direction $t$.
We will further distinguish two different subcases according to whether
the $Dp$-brane extends along the spatial non-compact direction $x$ (in which case it is spacetime-filling) or not (in which case it is a domain wall).

\subsubsection{Spacetime-filling $Dp$-branes}
Let us  assume that $\Sigma$ wraps $t$, $x$ and an odd $(p-1)$-cycle inside $\mcal_8$. Explicitly let $\Sigma$ be parameterized by coordinates $(t,x,\sigma^a)$, $a=1,\dots,p-1$, so that
\eq{
x^m=x^m(\sigma)
~,}
where $x^m$ are coordinates of $\mcal_8$. The condition (\ref{cor1}) implies that the worldvolume fieldstrength is of the form:
\eq{\label{313}\fcal_{\text{wv}}=-\hcal^{-1}\d t\wedge\d x+ \d x\wedge f+\hat{\fcal}_{\text{wv}}~,}
where we have defined
\eq{f=f_a~\!\d\sigma^a~;~~~\hat{\fcal}_{\text{wv}}=\frac{1}{2}\left(\hat{\fcal}_{\text{wv}}\right)_{ab}~\!\d\sigma^a\wedge\d\sigma^b~.}
Moreover it follows from the form of the NSNS three-form in (\ref{nsf}) that for the worldvolume Bianchi identity (\ref{wvbianchi}) to be satisfied,the last two terms on the right-hand side of (\ref{313})  must be closed. Note that the electric worldvolume field is necessarily  non-vanishing.

Taking (\ref{data}) into account, equations (\ref{nscal}) can be seen to reduce to the following condition
\eq{\label{reducedcalib}\left[\hcal^{-1/2}f\wedge\varphi\bs\wedge e^{\hat{\fcal}_{\text{wv}}}\right]_{p-1}
=\sqrt{-\det(\gs+\fcal_{\text{wv}})}~\!\d^{p-1}\!\sigma
~,}
where on the left-hand side above it is understood that we keep only the  $(p-1)$-form.

Let us now describe explicitly some calibrated spacetime-filling branes.

\subsubsection*{D2}

Consider the case of a D2 brane $\Sigma$ extending along $(t,x)$ and wrapping an internal direction parametrized by $\psi$. We take $\psi$ such that
\al{
\n|_\Sigma = - \d \psi \;, \quad \s_j|_\Sigma = \tilde{\s}_j|_\Sigma = 0
}
for the left-invariant forms\footnote{See \cite{kp} for an explicit parametrization of the left-invariant forms in terms of coordinates on $V_{5,2}$. \\
The left-invariant forms $\s_j$ are not to be confused with the coordinates $\s^a$ on $\Sigma$. We do not refer to $\s_j$ elsewhere in this section, and we do not refer to $\s^a$ outside of this section.}  $\n$, $\s_j$, $\tilde{\s}_j$ on $V_{5,2}$.

Specializing (\ref{313}) to the case at hand, the worldvolume fieldstrength reads:
\eq{\fcal_{\text{wv}}=-\hcal^{-1}\d t\wedge\d x+\d x\wedge f~;~~~f=f_{\psi}\d\psi~;~~~\hat{\fcal}_{\text{wv}}=0~,}
and automatically satisfies (\ref{wvbianchi}) for $f_{\psi}$ an arbitrary function of $x$, $\psi$. Moreover:
\eq{\left[f\wedge\varphi\bs\wedge e^{\hat{\fcal}_{\text{wv}}}\right]_{1}=f_{\psi}\d\psi~,}
and
\eq{\label{matrix1}
\gs+\fcal_{\text{wv}}
=\left(
\begin{array}{ccc}
-\hcal^{-1} & -\hcal^{-1} &0\\
\hcal^{-1} & \hcal^{-1} & f_{\psi}\\
0& -f_{\psi}& c^2\\
\end{array}
\right)
\Longrightarrow -\det(\gs+\fcal_{\text{wv}})=H^{-1}f_{\psi}^2
~,}
so that the calibration condition (\ref{reducedcalib}) is satisfied provided $f_{\psi}$ is everywhere positive. Note that the DBI part of the action evaluates to:
\eq{\label{dbid2}
S^{\mathrm{DBI}}_{D2}= -\mu_2\int_{\Sigma} e^{-\phi}
\sqrt{-\det(\gs+\fcal_{\text{wv}})}~\!\d^{3}\!\sigma
=-\mu_2\int_{\Sigma} \frac{1}{g_s}\hcal^{1/2}\hcal^{-1/2}f_{\psi}\d t\d x\d\psi
~.}
The above is in agreement with the BPS bound (\ref{dbi}), as it should.

Note that the fieldstrength $\fcal_{\text{wv}}$ is not closed, in accordance with (\ref{wvbianchi}). To identify the wotldvolume electromagnetic field $F$ in $\fcal_{\text{wv}}$ we should split:
\eq{\fcal_{\text{wv}}=B\bs+2\pi\alpha' F~,}
with $F$ closed. The form of $B$ will impose constraints on the quantization of Page charges in the RR sector.

We should finally check the worldvolume equations of motion. Consider (\ref{defp}), (\ref{defpi}). In order to calculate these explicitly, the RR flux needs to be specified explicitly. For convenience, we consider only massless IIA, i.e., we set the RR flux $\fcal_0 = F_0 = 0$. Then regardless of which other RR fluxes we turn on (other scalar terms, (1,1)-forms, (2,2)-forms as discussed in the previous section), the Wess-Zumino term reduces to
\al{
\left.C\wedge e^{\fcal_{wv}}\right|_{\Sigma} =0 \;.
}
This results in:
\eq{
P_M^{\beta}=\frac{1}{g_s}\left(
\begin{array}{ccc}
f_{\psi}&0&0\\
0&0&0\\
c^2&0&0\\
\end{array}
\right)~;~~~
\Pi^{\alpha\beta}=\frac{1}{g_s}\left(
\begin{array}{ccc}
0&-\frac{c^2}{f_{\psi}}&0\\
\frac{c^2}{f_{\psi}}&0&-1\\
0&1&0\\
\end{array}
\right)
~,}
where the indices $M$, $\alpha$ should be understood
as enumerating the rows, while the index $\beta$ enumerates the columns.
It can then be checked that the $X^M$-eoms (\ref{xeom}) are automatically satisfied. The $A_{\alpha}$-eoms (\ref{aeom}) are also automatically satisfied except for the Gauss law constraint $\partial_{\alpha}\Pi^{t\alpha}=0$, which imposes that $f_{\psi}$ should only depend on the coordinate $\psi$.

\subsubsection{Domain wall $Dp$-branes}
Let us now assume that $\Sigma$ wraps $t$ and an even $p$-cycle inside $\mcal_8$. Explicitly let $\Sigma$ be parameterized by coordinates $(t,\sigma^a)$, $a=1,\dots,p$, so that
\eq{
x^m=x^m(\sigma)
~,}
where $x^m$ are coordinates of $\mcal_8$. The condition (\ref{cor1}) implies that the worldvolume fieldstrength is of the form:
\eq{\fcal_{\text{wv}}=\frac{1}{2}\left(\fcal_{\text{wv}}\right)_{ab}~\!\d\sigma^a\wedge\d\sigma^b~,}
i.e. contrary to the spacetime-filling case here there can be no electric worldvolume field.
Moreover it follows from the form of the NSNS three-form in \eqref{nsf} that for the worldvolume Bianchi identity (\ref{wvbianchi}) to be satisfied  ${\fcal_{\text{wv}}}$ must be closed.

Taking (\ref{data}) into account, equations (\ref{nscal}) can be seen to reduce to the following condition
\eq{\label{reducedcalib2}\left[\hcal^{-1/2}\varphi\bs\wedge e^{\fcal_{\text{wv}}}\right]_p
=\sqrt{-\det(\gs+\fcal_{\text{wv}})}~\!\d^{p}\!\sigma
~,}
where on the left-hand side above it is understood that we keep only the  $p$-form.

\section{$SU(4)$-structure deformations of Stenzel space}\label{deforms}
So far, we have discussed supersymmetric vacua of type II and M-theory on Stenzel space, by applying the known results for eight-manifolds with $SU(4)$-structure: Stenzel space is a Calabi-Yau fourfold, hence it has such an $SU(4)$-structure, as determined by \eqref{eq:stenzelsu}. In order to go beyond Stenzel space, we will alter the geometry by considering deformations of the $SU(4)$-structure.
\\
\\
Our starting point will be to consider the topological space $\scal \equiv T^* S^4 $, which is homeomorphic to Stenzel space. We will first investigate all $SU(4)$-structures on this space which are `left-invariant'; we will explain what we mean by this. We will then consider a subset of these, which we dub `$abc$ $SU(4)$-structures'. The canonical metric of such $abc$ $SU(4)$-structures is identical to the Stenzel metric \eqref{eq:stenzelmetric}, but with $a,b,c$ generic functions of the radial direction $\tau$ rather than fixed to satisfy \eqref{eq:stenzelabc}.
\\
\\
By taking generic $a,b,c$, we find that $J, \O$ are no longer closed and thus torsion classes have been turned on, going beyond the Calabi-Yau scenario. As the torsion classes determine integrability of geometrical structures associated to the $SU(4)$-structure (such as the almost complex and almost symplectic structure),  we can examine moduli spaces of such geometrical structures. These moduli spaces are  determined by differential equations which we can solve for a number of cases.
\\
\\
Finally, we should take care that by altering the metric, we do not make the space geodesically incomplete. We analyze various possibilities, and conclude that to avoid this, the straightforward thing to do is to ensure that at $\tau =0$, the space maintains its $S^4$ bolt, albeit possibly squashed.

\subsection{Left-invariant $SU(4)$-structures}
As mentioned, our starting point is $\scal = T^* S^4$  equipped with a Riemannian metric $g$. In order to construct $SU(4)$-structures on this space, we will require some knowledge of the cotangent bundle of $\scal$. Due to its conical structure, we will first focus on the forms on $V_{5,2}$, and for that, it is convenient to first discuss left-invariant forms on cosets.
We consider the left-invariant forms of $SO(5)$, descended on the coset $V_{5,2} \simeq SO(5) / SO(3)$. Concretely, this means the following: the left-invariant one-forms of $SO(5)$ are given by $\n, \s_j, \tilde{\s}_j, L_{jk}$, $j,k \in \{1,2,3\}$, with $L_{jk}$ the left-invariant one-forms of the $SO(3)$ subgroup, while $\s_j, \tilde{\s}_j, \n$ lie in the complement. In terms of these, a $p$-form on $V_{5,2}$ is left-invariant if and only if its exterior derivative lies in the complement, i.e., is expressible solely in terms of $\s_j, \tilde{\s}_j, \n$. Given a left-invariant form, any scalar multiple is also a left-invariant form. The exterior derivative maps left-invariant $p$-forms to left-invariant $p+1$-forms. For more general details, see \cite{nieuw}.
\\
\\
A basis of left-invariant forms up to fourth degree for $V_{5,2}$ and their derivatives is given in the following table (see also \cite{varela}):
\begin{center}
\begin{tabular}{ |l|l|l|}
\hline
1-forms & $\n$ &  \\\hline
2-forms & $d\n = \tilde{\s}^j \wedge \s^j$ &\\\hline
\multirow{5}{*}{3-forms}
& $\a_0=\n \wedge d\n$ &  $\d \a_0 = \b_0$ \\
& $\a_1=\s^1 \wedge \s^2 \wedge \s^3$ & $\d\a_1 = - \b_4$  \\
& $\a_2=\tilde{\s}^1 \wedge \tilde{\s}^2 \wedge \tilde{\s}^3$   & $\d \a_2 = - \b_3$ \\
& $\a_3=\frac{1}{2} \varepsilon_{ijk} \tilde{\s}^i \wedge \s^j \wedge \s^k$ & $\d \a_3 = 2 \b_3 - 3\b_2$ \\
& $\a_4= \frac{1}{2}\varepsilon_{ijk} \s^i \wedge \tilde{\s}^j \wedge \tilde{\s}^k$  & $\d\a_4 = 2\b_4 - 3 \b_1$  \\ \hline
\multirow{5}{*}{4-forms}
& $\b_0=d\n \wedge d\n$ &\\
& $\b_1= - \n \wedge \a_2$ &\\
& $\b_2=\n \wedge \a_1$ &\\
& $\b_3= \n \wedge \a_4 $ &\\
& $\b_4=- \n \wedge \a_3$ &\\ \hline
\end{tabular}\label{li-forms}
\end{center}
Although left-invariant forms only make sense given a specific group action on the space of forms, we will abuse terminology and make the following definition: Let $\tau$ be the radial coordinate on $\scal$ viewed as a (deformed) cone over $V_{5,2}$. Then we define a \textit{left-invariant form on $\scal$} to be a form that restricts to a left-invariant (`LI') form on $V_{5,2}$ at any fixed $\tau$. This is equivalent to demanding that the exterior derivative acting on such forms can be expressed purely in terms of the radial one-form $\d\tau$ and the left-invariant one-forms of $SO(5)$ that lie in the coset $SO(5)/SO(3)$, i.e., $\n, \s_j, \tilde{\s}_j$.
All such LI $p$-forms on $\scal$ can be gotten by taking linear combinations of LI $p$-forms on $V_{5,2}$, and LI $p-1$-forms on $V_{5,2}$ wedged with $\d\tau$, with coefficients that are functions only of $\tau$.

We define a \textit{left-invariant $SU(4)$-structure} to be an $SU(4)$-structure defined by forms $(J, \O)$ such that $J$ and $\O$ are left-invariant forms on $\scal$\footnote{In\cite{varela}, the LI forms on $V_{5,2}$ are used to construct LI $SU(3)$-structures on $V_{5.2}$.}. The most general LI two-form and LI four-form are given by
\all{
J =& n (\tau) \n \wedge \d\tau - m (\tau) d\nu\\
\O =& k_j(\tau)   \d\tau \wedge \a^j + l_j (\tau) \b^j \;,
}
with $j \in \{ 0, ..., 4\}$ summed over. In order for $(J, \O)$ to form an $SU(4)$-structure, it needs to satisfy the following constraints:
\eq{\spl{\label{eq:su4cons}
J\wedge \O =& 0\\
\frac{1}{4!}J^4 =& \text{vol}_8\\
\frac{1}{2^4} \O \wedge \O^* =& \text{vol}_8\\
\star \O =& \O
\;.
}}
In addition, $\O$ needs to be decomposable\footnote{
In $d=6$, this can be checked by means of the Hitchin functional \cite{hitchin}: The three-form $\O$ determines the complex structure through
\all{
I_j^{\phantom{j}k} = - \epsilon_{j m_1 ... m_5 } \left(\text{Re}\O\right)^{k m_1 m_2} \left(\text{Re}\O\right)^{m_3 m_4 m_5}\;.
}
In $d=8$, this fails, as one can see from a selfconsistency check: one finds instead that
\all{
- \epsilon_{j m_1 ... m_7 } \left(\text{Re}\O\right)^{k m_1 m_2 m_3} \left(\text{Re}\O\right)^{m_4 m_5 m_6 m_7} = \delta_j^k \;.
}
This is a consequence of the fact that $\star_8\O$ = $\O$ compared with $\star_6 \O = i \O$.} and holomorphic. Note that the final constraint is not independent, yet explicitly working with this redundancy is convenient. Let us see how these constrain the free parameters $k_j$, $l_j$, $n$, $m$. We start by introducing normalized one-forms
\eq{\begin{alignedat}{2}
e_0 &= \frac{1}{\sqrt{g^*(\d\tau, \d\tau)}} \d\tau \;, \quad &e_j = \frac{1}{\sqrt{g^*(\s_j, \s_j)}} \s_j \phantom{\;,}\\
\tilde{e}_0 &= \frac{1}{\sqrt{g^*(\n, \n)}} \n \;, \quad &\tilde{e}_j= \frac{1}{\sqrt{g^*(\tilde{\s}_j, \tilde{\s}_j)}} \tilde{\s}_j\;,
\end{alignedat}}
with $g^*$ the induced metric on the cotangent bundle. We shift the coefficients of $J, \O$ to encompass these normalizations so that we can consider the one-forms normalized without loss of generality.
The first constraint of \eqref{eq:su4cons} sets $k_0 = l_0= 0$.
Let us now choose an orientation by taking
\eq{\spl{
\text{vol}_8 &= \sqrt{g} d\tau \wedge \n \wedge \s_1 \wedge \s_2 \wedge \s_3 \wedge \tilde{\s}_1\wedge \tilde{\s}_3\wedge \tilde{\s}_3\\
&= e_0 \wedge  \tilde{e}_0 \wedge  e_1\wedge  e_2\wedge  e_3\wedge  \tilde{e}_1\wedge  \tilde{e}_2\wedge  \tilde{e}_3\;.
}}
Then the selfduality constraint leads to $k_j = l_j$.
Next, we consider decomposability of $\O$. As detailed in \cite{ckkltz} appendix C, a $p$-form $\o$ is decomposable iff $\forall X \in \mathfrak{X}^{p-1}(M)$, $\iota_X \o \wedge \o = 0$. This gives us 56 equations, a number of which are trivial while the rest are linearly dependent. They boil down to
\all{
k_1^2 + k_2^2 =& 0 \\
k_3^2 + k_4^2 =& 0 \\
k_1^2 + k_3^2 =& 0 \\
k_2^2 + k_4^2 =& 0 \\
k_3 ( k_1 + k_4) =& 0\\
k_4 (k_2 + k_3 )=& 0\;,
}
with solution
\al{
\vec{k} = (k_1, \pm i k_1, \mp i k_1, -k_1)\;.
}
As we will see, the choice of $\pm$ determines which one-forms are holomorphic and which are anti-holomorphic.
Using this solution, the volume constraints reduce to
\all{
1 =& nm^3 = |k_1|^2\;,
}
hence $k_1$ is just a phase. Holomorphic one-forms are then given by
\eq{\spl{
\zeta^0 =& e^{ i \varphi_0} \left(- e_0 + i \tilde{e_0}\right)\\
\zeta^j =& e^{ i \varphi_j} \left( e_j + i \tilde{e_j}\right)\;,
}}
with $e^{ i (\varphi_0+ \varphi_1 + \varphi_2 +\varphi_3) }= k_1$ such that
\al{
\O = \zeta^0 \wedge \zeta^1  \wedge \zeta^2  \wedge \zeta^3\;.
}
In terms of these holomorphic one-forms, we have
\al{
J = \frac{i}{2} \left( n \z^0 \wedge \bar{\z}^0 + m \z^j \wedge \bar{\z}^j \right)\;.
}
Since the complex structure is given by $I_a^{\phantom{a}b}= J_{ac}g^{cb}$, $\O$ is holomorphic, and $g^*(\z^\a, \bar{\z}^\a) = 2$, the constraint $I^2 = - \mathbbm{1}$ enforces
$n = m = 1$, leading to
\al{
J = \frac{i}{2} \zeta^\a \wedge \bar{\zeta}^\a
}
and the metric must be given by
\eq{\spl{
ds^2(\scal) =&\frac{1}{2} \left( \zeta^\a \otimes \bar{\zeta}^\a + \bar{\zeta}^\a \otimes \zeta^\a \right) \\
=& e^\a \otimes e^\a + \tilde{e}^\a \otimes \tilde{e}^\a \\
\equiv& \frac{c^2}{4} \d\tau^2 + c^2 \n^2 + a_j \s_j^2 + b_j \tilde{\s}_j^2\;,
}}
with the additional restriction
\al{
a_1 b_1 = a_2 b_2 = a_3 b_3 \;.
}
Here, we have rescaled $\tau$ such that its coefficient is once more $\frac{c^2}{4}$ and renamed the normalization to match the notation in the Stenzel space scenario. The rotations generated by the four angles $\varphi_0, \varphi_j$ leave the $SU(4)$-structure invariant and hence can be set to zero without loss of generality. Thus, LI $SU(4)$-structures are defined by the five free parameters  $a_1(\tau), b_1(\tau), b_2(\tau), b_3 (\tau), c(\tau)$.

\subsection{$abc$ $SU(4)$-structures}
We will restrict ourselves to the case $a_j = a$, $b_j = b$ $\forall j$, as the generalization does not lead to any novel features outside of more cluttered equations. For lack of imagination, we shall refer to such as $abc$ $SU(4)$-structures to distinguish them from the more general LI $SU(4)$-structures.
Let us explicitly spell out such structures. We have orthonormal one-forms
\eq{
\begin{alignedat}{2}
e_0 =& \frac{c(\tau)}{2} \d\tau \;, \quad &e_j = a(\tau) \s_j \\
\tilde{e}_0 =& c(\tau) \n \;, \quad &\tilde{e}_j= b(\tau) \tilde{\s}_j
\end{alignedat}
}
and holomorphic one-forms
\eq{\spl{
\z^0 &= - \frac{c}{2}d\tau +i c \n \\
\z^j &= a \s_j + i b \tilde{\s}_j \;.
}}
The metric and volume form are given by
\al{
ds^2(\scal) =& (e^\a)^2 + (\tilde{e}^\a)^2 \nn \\
\text{vol}_8 =& e^0 \wedge \tilde{e}^0 \wedge e^1 \wedge e^2 \wedge e^3 \wedge \tilde{e}^1 \wedge \tilde{e}^2 \wedge \tilde{e}^3   \;,
}
with $\a \in \{0,1,2,3\}$, and the $SU(4)$-structure is given by
\eq{\spl{\label{j}
J &= \frac{i}{2}\z^\a \wedge \bar{\z}^{\bar{\a}}\\
\O &= \z^0 \wedge \z^1 \wedge \z^2 \wedge \z^3
~.}}
For $a,b,c$ as in \eqref{eq:stenzelabc}, this $SU(4)$-structure is the one on Stenzel space.
As explained in section \ref{SU4}, the $SU(4)$-structure determines the existence of geometrical structure in terms of torsion classes, defined by
\eq{\spl{
\d J =& W_1 \lrcorner \O^* + W_3 + W_4 \wedge J + \text{c.c.} \\
\d \O =& \frac{8 i}{3} W_1 \wedge J \wedge J + W_2 \wedge J + W_5^* \wedge \O \;.
}}
Using the explicit expressions for $J$ and $\O$, one finds that
\al{\label{eq:torsion}
W_1 &= 0 \nn \\
W_2 &=\frac{1}{4i abc}(a^2 - b^2 + 2 b a'-2 a b') \varepsilon_{\bar{i}jk} \bar{\z}_i \wedge \z_j \wedge \z_k \nn\\
W_3 &= 0 \\
W_4 &= - \frac{1}{abc}\left((ab)'- \frac{1}{2}c^2\right)\z_0\nn\\
W_5 &=-\frac{1}{4 abc^2}\left(-3 (a^2+  b^2) c + 6 (ab)'c +  4 abc'\right)\z_0\;.\nn
}
Note that these equations are only sensible as long as $a,b,c \neq0$. At any point where this does not hold, the $SU(4)$-structure degenerates: as an example, let $p \in \scal$ be described in local coordinates as  $p = (\tau_0,...)$ and let $b$ satisfy  $b(\tau_0) = 0$. Then the metric reduces to
\all{
\left.g\right|_{T_p\scal}  =  c(\tau_0)^2 \left( \frac14 \d \tau^2 + \n^2 \right) +  a(\tau_0)^2 \s_j^2\;.
}
Let the vectors $\tilde{v}_j \in T_p \scal$ be dual to $\left.\tilde{\s}_j\right|_p$. Then
\all{
\left.g\right|_{T_p\scal} (\tilde{v}_j, v) = 0 \quad \forall v \in T_p \scal
}
and thus the metric is degenerate. On Stenzel space, such a situation occurs at the tip, where $b(0) = 0$  leads to an $S^4$ bolt. In particular, this $S^4$ bolt is a special Lagrangian submanifold with $J=0$.

\subsubsection{Moduli spaces}\label{moduli}
Let us examine the geometry determined by the torsion classes more carefully.
As a consequence of \eqref{eq:torsion}, we see that whatever functions we choose for $a,b,c$, $W_1 = W_3 = 0$. This leaves us with $W_2, W_4, W_5$ which can all either be zero or non-zero. Let us first consider $W_2$. The almost complex structure determined by the $SU(4)$-structure is integrable if and only if $W_2=0$, which is equivalent to
\al{\label{eq:W2}
a^2 - b^2 + 2 b a'-2 a b' = 0\;.
}
By substituting $r = b/a$ one can find the solution to this equation to be given by
\al{
r \in \left\{\pm 1, \frac{e^{(\tau + \tau_0)/2} \mp  e^{- (\tau+ \tau_0)/2} }{e^{(\tau+ \tau_0) /2} \pm  e^{- (\tau+ \tau_0)/2}}\;|\; \tau_0 \in \mathbbm{R} \right\}
}
where one should consider $r = \pm1$ to be the limiting cases for the integration constant $\tau_0 \rightarrow \pm \infty$. The moduli space of complex structures for $SU(4)$-structures satisfying our ansatz is then given by these solutions, modulo diffeomorphisms (in particular, $\tau \rightarrow \tau - \tau_0$ and $\tau \rightarrow - \tau - \tau_0$):
\al{
\mathfrak{M}_\mathbbm{C}&= \left\{(a,r a,c)\;|\; r \in \{ \pm 1, \tanh^{\pm1} \left(\frac{\tau}{2}\right)\}  \right\}\;.
}
Note that  $ r \rightarrow r^{-1}$ leaves the moduli space invariant, which is a consequence of  \eqref{eq:W2} being invariant under $a \leftrightarrow b$. In essence, we thus see that there are but two possible complex structures: we will refer to $r = \tanh \left(\frac{\tau}{2}\right)$ as the Stenzel complex structure, and to $r =  1 $ as the conical complex structure (we will describe this in more detail in the next subsection).

Next, let us examine when the $SU(4)$-structure determines an integrable almost symplectic structure. The non-degenerate two-form $J$ is closed (and hence, a symplectic form) if and only if $W_4=0$, which is equivalent to
\al{
(ab)'- \frac{1}{2}c^2 = 0\;,
}
hence the moduli space of symplectic structures is given by
\al{
\mathfrak{M}_S &=\left\{(a,b, \sqrt{2(ab)'} ) \right\}
}
Compatibility of the complex and symplectic structure determines a K\"{a}hler structure, hence the K\"{a}hler moduli space is given by
\al{
\mathfrak{M}_K &= \mathfrak{M}_\mathbbm{C} \cap \mathfrak{M}_S \nn\\
&= \left\{(a,r a,\sqrt{2(r a^2)'})\;|\; r \in \{ \pm 1, \tanh^{\pm1} \left(\frac{\tau}{2}\right)\}  \right\}
}
Conformal Calabi-Yau structures are found by demanding $W_2 = 0$, $2W_4 = W_5$. Setting $b = r a$ and using the expressions for $W_4, W_5$, this reduces to the equation
\al{\label{eq:ccy}
\tilde{r}' + \left(r + 2 r^{-1} \right) \tilde{r} - 2 r^{-1} = 0\;,
}
where we have set $\tilde{r} \equiv a^2 / c^2$. This is solved by
\al{
\tilde{r} = \left\{\label{eq:ccy2}
\renewcommand\arraystretch{2}
\begin{array}{cl}
 \frac{2 + \cosh\left(\tau\right) + k \left(\sinh\left(\frac{\tau}{2}\right)\right)^{-4} }{3 \cosh^2\left(\frac{\tau}{2}\right)} & \phantom{abcde} r = \tanh \left(\frac{\tau}{2}\right) \\
\frac{2}{3} + k e^{\mp 3 \tau} &  \phantom{abcde} r = \pm1 \\
\end{array} \right.
}
for $k \in \mathbbm{R}$. As $(r, \tilde{r})$ determine the ratios between $a$ and $c$ and between $a$ and $b$, and one can by conformal transformation set $a=1$, this determines every possible CCY-structure. Some remarks are in order here. Firstly, note that if we consider $\tilde{r}$ as a function of $r$,
\all{
\lim_{\tau \rightarrow \pm \infty} \tilde{r}\left(\tanh \left(\frac{\tau}{2}\right)\right) = \tilde{r}( \pm 1)
}
after rescaling $k$, thus confirming our earlier argument that such points in the moduli space of complex structures should be considered as limiting cases.
Secondly,  note that for $r = \tanh \left(\frac{\tau}{2}\right)$,  $\tilde{r}$ is singular at $\tau = 0$ indicating that the $SU(4)$-structure degenerates as either $c$ approaches $0$ or $a$ blows up. This happens unless one sets the integration constant $k = 0$, which gives the unique regular solution
\al{\label{eq:stenzelccy}
\tilde{r} =  \frac{2 + \cosh\left(\tau\right) }{3 \cosh^2\left(\frac{\tau}{2}\right)} \;.
}
This is exactly the proportionality between $a^2$ and $c^2$ for Stenzel space \eqref{eq:stenzelabc}, so we have found that the only smooth CCY manifolds with the Stenzel space complex structure are in fact conformal to Stenzel space. On the other hand, the conical complex structures lead to a regular $\tilde{r}$ regardless of choice of $k$. Thirdly, note that $r = 1/ \tanh \left(\frac{\tau}{2}\right)$ is uninteresting as the constraint  $2W_4 = W_5$ is invariant under $a \leftrightarrow b$. Thus the solution is as above but with $a \leftrightarrow b$.

Finally, a Calabi-Yau structure is found by demanding $W_2 = W_4 = W_5 = 0$. Note that this is equivalent to \eqref{eq:cycondition} as given in \cite{cglp}.
In case we take $r=  \tanh \left(\frac{\tau}{2}\right)$, we can solve $W_5 = 0$ to find
\all{
r a^2 =  \frac{1}{2} \l^{8/3}  c^{-2/3}\sinh \tau \;, \qquad \l \in \mathbbm{R}
}
and use this to solve $W_4 =0$, leading to
\all{
c^2 &= 3^\frac{3}{4}    \l^2 \frac{\cosh^3\left(\frac{\tau}{2}\right)}{\left(\frac{k}{\sinh^4\left(\frac{\tau}{2}\right)} + \left(2 + \cosh \tau \right)\right)^{3/4}}\\
a^2 &= 3^{-\frac{1}{4}} \l^2 \cosh\left(\frac{\tau}{2}\right)                                   \left(\frac{k}{\sinh^4\left(\frac{\tau}{2}\right)} + \left(2 + \cosh \tau \right)\right)^{1/4}\\
b^2 &= 3^{-\frac{1}{4}} \l^2 \cosh\left(\frac{\tau}{2}\right)\tanh^2\left(\frac{\tau}{2}\right) \left(\frac{k}{\sinh^4\left(\frac{\tau}{2}\right)} + \left(2 + \cosh \tau \right)\right)^{1/4}
}
Compare with \eqref{eq:stenzelabc}. This derivation is completely equivalent to the derivation of the Stenzel metric in \cite{cglp}, except we have explicitly kept an integration constant $k$: however, any non-zero $k$ will lead to a singularity at $\tau = 0$. For our intents and purposes, we are thus only interested in Stenzel space.

In case we take $r = 1$, the solution to $W_5=0$ is given by
\all{
a^2 = \frac{2}{3} \l^{8/3} c^{-2/3} e^\tau
}
and solving $W_4$ then leads to
\eq{\spl{\label{eq:cone2}
c^2 &=\l^2 \frac{e^{3\tau}}{(e^{3\tau}+ k)^{3/4}} \\
a^2 &= b^2 = \frac{2}{3} \l^2 (e^{3\tau}+ k)^{1/4}
}}
If instead we take $r= -1$, the same solution is found but with $\tau \rightarrow - \tau$ as expected.

\subsubsection{Geodesical Completeness \& Cones} \label{cones}
Before we start constructing vacua using these $abc$ $SU(4)$-structures, let us first consider the construction of CY-structures with $r=1$, i.e., \eqref{eq:cone2}.
In order to illustrate what is happening here we consider the specific case $k=0$, such that this configuration reduces to
\eq{\spl{\label{eq:cone}
a &=  b = \sqrt{\frac{2}{3}} c\\
c &= \l e^{ \frac38 \tau} \;.
}}
In the UV, this solution is identical to Stenzel space. In the IR it behaves quite differently; at $\tau=0$, $b \neq 0$ hence there is no $S^4$ bolt. Indeed, as $a$, $c$ also do not vanish, $\tau=0$ is no special point and we see that this slice is just a copy of $V_{5,2}$, just like slices for any other $\tau$. In fact, we can make a coordinate transformation to find that the metric is globally given by
\al{
ds^2(\scal) &= \l^2 e^{\frac{3}{4} \tau} \left( \frac{1}{4} \d \tau^2 + \n^2 + \frac{2}{3} (\s_j^2 + \tilde{\s}_j^2 )\right)\nn\\
&=  \frac{16}{9} \l^2  \left[\d R^2 +  R^2 ds^2(V_{5,2}) \right]
}
with $ds^2(V_{5,2})$ defined in \eqref{eq:confadspara} and $R \in [1,\infty)$ for $\tau \in [0, \infty)$. We recognize that this is just the undeformed cone with the tip `cut off', as it were. More precisely, the space is geodesically incomplete: one can solve the geodesic equation to find
\all{
\tau(t) = k_2 - \frac{8}{3} \log(3t + k_1)
}
which is not a solution $\forall t \in \mathbbm{R}$, regardless of $k_{1,2}$. Obviously, the tip of the cone is a singularity; smoothing out this singularity was the reason Stenzel space garnered interest in the first place.
\\
\\
We illustrate this because it points us to a potential pitfall: one can consider arbitrary $abc$ $SU(4)$-structures, but without the $S^4$ bolt at the tip it is not guaranteed that the space is geodesically complete. We have the following possibilities at $\tau=0$:
\begin{itemize}
\item $a=b=c=0$ leads to a singularity. $a,b,c \neq 0$ leads to potentially incomplete spaces as above; all spaces of such type that we have examined are incomplete and have a conical singularity in their completion.
\item $a=0$, $b=  c \neq 0$ leads to an $S^4$ bolt, similar to $b=0$, $a= c \neq 0$, on which $\O$ is proportional to the volume form and $J$ vanishes. This can be deduced by noting that the defining equation for the LI forms of $V_{5,2}$ \eqref{eq:liforms} are invariant under $\n \rightarrow - \nu$, $\s_j \leftrightarrow \tilde{\s}_j$. This transformation interchanges the four-forms \eqref{eq:(2,2)NL1} and \eqref{eq:(2,2)L1}, which also explains why $ a\leftrightarrow b$ leads to $f_L \leftrightarrow f_{NL}$ as follows from \eqref{eq:(2,2)NL2}, \eqref{eq:(2,2)L2}. Thus, which (2,2)-form diverges and which does not is interchanged. This conclusion essentially remains the same for vacua on $SU(4)$-deformed spaces. In case one has $b=0$, $a, c \neq 0$, $a \neq c$ (or similarly for $a \leftrightarrow b$), the bolt will be a squashed $S^4$, with squashing parameter $c(0) / a(0) \equiv \a(0)$, as follows from the metric at $\tau = 0$.
\item The remaining possibilities are
\begin{itemize}
\item[-]$ c = 0, a, b \neq 0$
\item[-] $a = c = 0, b\neq 0$ or $b = c = 0, a\neq 0$
\item[-]$a=b=0$, $c \neq 0$.
\end{itemize}
For these cases, it is unclear whether or not the space is singular at $\tau =0$ (i.e., whether or not the curvature blows up); this is purely due to computational difficulties, as one should be able to calculate the Riemann tensor to see whether or not this is the case.
\end{itemize}
Due to these considerations, we will limit ourselves to cases where the metric comes with a (possibly squashed) $S^4$ bolt to ensure that we do not encounter potential incompleteness issues. As both types of $S^4$ bolts are similar up to transformations $a \leftrightarrow b$, we will only consider the case where
\al{\label{eq:geo}
a(0), c(0) \neq 0 \;,\quad b(0) = 0\;.
}
One can consider this a boundary condition on the differential equations determining our vacua.

\section{Type IIA theory on $SU(4)$-structure deformed Stenzel Space}\label{generalizations}
We have discussed supersymmetric vacua of type IIA supergravity and M-theory on Stenzel space, and $SU(4)$-structure deformations of Stenzel space. In this section, we will discuss $\ncal= (1,1)$ vacua of type IIA supergravity on  on $SU(4)$-structure deformed Stenzel space. Our parameters consist of $a,b,c$, which determine the $SU(4)$-structure, and the RR fluxes. We would like to choose these in such a way that the following hold:
\begin{itemize}
\item[1)] $a,b,c$ are such that that $\scal$ equipped with such an $abc$ $SU(4)$-structure is geodesically complete and free of singularities.
\item[2)] The torsion classes and fluxes satisfy one of the branches of the IIA susy solutions, as given in \ref{typeII}.
\item[3)] The Bianchi identities are satisfied.
\item[4)] The integrability conditions, which turn a solution to the susy equations into a solution of the equations of motion, are satisfied.
\item[5)] The warp factor is regular and positive.
\item[6)] All the fluxes are $L^2$.
\end{itemize}
Unfortunately, due to the particularities of left-invariant forms on $\scal$, one cannot go beyond the Stenzel space scenario satisfying the first three constraints. A way out is by discarding the constraint that the Bianchi identities are satisfied. Such violations can come about in the presence of sources, which modify the action. The precise source terms needed can be deduced from the integrability equations. More precisely, we will see that the NSNS Bianchi identity will always be violated, thus indicating the presence of NS5-branes. On the other hand, the RR Bianchi identities need not be violated.
\\
\\
In the rest of this section, we will demonstrate the claims in the above paragraph. We then discuss vacua\footnote{There are, however, some subtleties to the integration theorem which we ignore. This will be explained in section \ref{sources}.} on $\scal$ which are complex but not symplectic (and hence, not CY), with primitive $(2,2)$-flux, with $\ncal=(1,1)$ supersymmetry, with external metrics that are asymptotically conformal $\text{AdS}_3$. We also give the appropriate source term. These vacua will not uplift to M-theory vacua on $\mathbbm{R}^{1,2} \times \scal$, as can readily be deduced from the torsion classes. In particular, these vacua will have the following properties:
\begin{itemize}
\item[1)] The metric will have an $S^4$ bolt at the origin and conical asymptotics, similar to Stenzel space, thus ensuring geodesical completeness.
\item[2)] The torsion class constraint imposed by supersymmetry can be explicitly solved, fixing $c$.
\item[3)] We consider the Bianchi identities on non-CY manifolds and deduce that taking solely a primitive (2,2)-form does not violate the RR Bianchi identities, whereas a number of other possibilities do. In all cases, $\d H \neq 0$.
\item[4)] We will explicitly check the integrability conditions in the presence of source terms. These yield the constraint which determines the source term needed.
\item[5)] The warp factor will be regular and positive, the flux $L^2$.
\end{itemize}

\subsection{A no-go for sourceless IIA $\ncal = (1,1)$ on $\scal$ with non-CY $abc$ $SU(4)$-structure}
We start by examining the constraints on the torsion classes imposed by $\mathcal{N}= (1,1)$ supersymmetry. As given by \eqref{eq:iiasol1}, \eqref{eq:iiasol2}, \eqref{eq:iiasol3},  $W_1 = W_3 = 0$ implies $W_2 = W_4 = W_5=0$ for $e^{2 i \t} \neq -1$. On the other hand, $e^{2 i \t} = -1$ allows for non-CY vacua. The remaining torsion classes are then related to $H$ as
\eq{\spl{ \label{eq:susycons}
\tilde{h}^{(1,0)}_3 =& \frac 14 \p^+ ( A-\phi)\\
W_2 =& - 2 i h^{(2,1)} \\
W_4 =& \p^+ ( \phi - A) \\
W_5 =& \frac32 \p^+ ( \phi - A) \;.
}}
Specifically, this implies that there are obstructions to the existence of a complex or symplectic structure if and only if the NS three-form has an internal component. That is to say, we must have that
\eq{\spl{ \label{eq:h3}
H_{3} =& - \frac14 W_4^* \lrcorner \O + \frac12 i W_2 + \text{c.c.} \\
\equiv& f(\tau) \z^1 \wedge \z^2 \wedge \z^3 + g(\tau) \varepsilon_{\bar{\imath}jk} \bar{\z}^{\bar{\imath}} \wedge \z^j \wedge \z^k+ \text{c.c.} \;,
}}
for some $f, g$ determined by $a,b,c$. Recall that $H$ is defined by \eqref{eq:hdecomp} and that \eqref{eq:hint} ensures that $\d H = 0$ if and only $\d H_3 = 0$.
It is easy to show that no functions $f,g$ exist that are non-trivial and are such that the Bianchi identity
\all{
\d H = 0
}
is satisfied. In fact, $H_3$ is a primitive left-invariant three-form, and there are no closed primitive left-invariant three forms as can be checked explicitly by using the basis for left-invariant forms on $V_{5,2}$ given in \ref{li-forms}. Thus, these classes of $SU(4)$-structures admit no non-CY vacua satisfying both susy and the Bianchi identities.
\\
\\
Now let us consider possible Calabi-Yau structures. We impose \eqref{eq:geo} to get a bolt at the origin, which means that we can pick only one possible complex structure, namely
\al{
r \equiv \frac{b}{a} = \tanh\left(\frac{\tau}{2}\right)\;.
}
As discussed in section \ref{moduli}, this complex structure allows only for Stenzel space as regular CY-structure.
\\
\\
Let us make some more remarks. First of all, for generic LI $SU(4)$-structures one still has $W_1 = W_3 = 0$.
Thus, the entire argument above goes through, and it can be concluded that there are also no non-CY vacua with LI $SU(4)$-structure satisfying susy and the Bianchi identities simultaneously. This is the primary reason we consider the more generic LI $SU(4)$-structures to be of little more interest than $abc$ $SU(4)$-structures.
Secondly, let us note that for any supersymmetric solution satisfying \eqref{eq:susycons} with non-vanishing torsion classes, one cannot set $e^\phi = g_s e^A$. As this is necessary to lift $d=2$ IIA vacua to M-theory vacua on $\mathbbm{R}^{1,2} \times \scal$, we immediately find that such IIA solutions do not uplift to M-theory vacua with external spacetime $\mathbbm{R}^{1,2}$. This is also evident from the fact that our $\ncal= 1$ M-theory vacua on $\mathbbm{R}^{1,2} \times \scal$ given in section \ref{m-theory} require \eqref{eq:mtorsion}, which immediately rules out non-CY vacua in the current case where $W_1 = W_3 = 0$.

\subsection{$\ncal = (1,1)$ IIA  torsion class constraint}
To summarize the previous section, there are no interesting non-Stenzel CY solutions, and non-CY solutions require violating the NSNS Bianchi identity. Accepting this for the moment, we consider the susy branch  $e^{2 i \t} = -1$, which, as noted, is the only branch that allows for non-CY with $W_1 = W_3 =0$.
\\
\\
Susy implies \eqref{eq:susycons}, which we take as defining equations for $H$, $\phi$. This constrains $a,b,c$ to satisfy
\al{
W_5 = \frac32 W_4 \;.
}
Plugging in \eqref{eq:torsion}
leads to the constraint
\al{\label{eq:iiasusycons}
- 3 ( a^2 + b^2 - c^2)c + 4 a b c' =0 \;.
}
As an aside, notice that this constraint is actually equivalent to the third equation of \eqref{eq:cycondition}.
One can solve this equation by parametrizing
\eq{\spl{\label{eq:para}
a \equiv& \a^{-1}(\tau) c\\
b \equiv& \b (\tau) a \;,
}}
leading to the solution
\al{\label{eq:susysol}
c (\tau) =& \l \exp\left(\frac34 \int^\tau dt \frac{1+ \b^2(t) - \a^2 (t)}{\b(t)} \right)   \;.
}
Generically, the space satisfies susy and is complex iff
\al{
\b = r \;.
}
It satisfies susy and is symplectic (and is in fact `nearly CY', i.e., only $W_2 \neq 0$) iff
\all{
3 \b \a' - \b'\a + 2 \a^5 - \frac32 \b^2 \a^3 - \frac 32 \a^3 =0 \;.
}

\subsection{RR Bianchi identities on non-CY manifolds}\label{rrbianchi}
The next constraints that we will examine are the RR Bianchi identities for IIA vacua which are not Calabi-Yau.
Generically, the RR Bianchi identities are given by
\eq{
\d_H \fcal = 0
}
In the CY case, they boil down to closure and co-closure of forms, see \eqref{eq:cybianchi}. However, for non-CY vacua, one instead finds that they reduce to
\eq{\label{eq:noncybianchi}
\d_{H_3} F = 0 \;.
}
Here, use has been made of the twisted selfduality condition \eqref{eq:11fluxes1} and the second line of \eqref{eq:hint}. Thus, in this case, the RR Bianchi identities no longer decouple to independent equations for each flux separately. We have four terms that can be turned on or off: $f_0$ \& $f_4$, $f_2$ \& $\tilde{f}_4$, $f^{(1,1)}$ and $f^{(2,2)}$. Thus, there are $2^4$ possibilities. We have examined a number of these:
\begin{itemize}
\item Turning on $f_0$ leads to
\all{
f_0 H_3 = \d( - F_2)\;.
}
As we already concluded in our examination of the NSNS Bianchi identity, there are no non-trivial primitive closed LI three-forms, and thus certainly no non-trivial primitive exact LI three-forms. Thus any configuration with $f_0 \neq 0$ violates the RR Bianchi identity.
\item Turning on solely $f_2$ \& $\tilde{f}_4$ or $f^{(1,1)}$ violates the RR Bianchi identity for any choice of $a,b,c$.
\item Turning on solely $f^{(2,2)}$, the RR Bianchi identity is satisfied if and only if $f^{(2,2)}$ is closed and co-closed and $W_2= 0$. Thus, we find similar constraints to the Stenzel case scenario: \eqref{eq:(2,2)NL2} for one and \eqref{eq:(2,2)L2} for the other primitive $(2,2)$-form. Note that these equations already demand $W_2 = 0$.
\end{itemize}
In cases where one turns on multiple (non-$f_0$) fluxes, there is a possibility that one could find configurations with $\a, \b$ fine-tuned in such a way that the Bianchi identities are satisfied (preferably with enough freedom to spare to satisfy all other demands). We have not been able to find these.

\subsection{Vacua on complex manifolds with $f^{(2,2)}$}\label{cplx22}
We consider vacua where the flux is given by a four-form RR flux, combined with the NSNS flux. The four-form is primitive, (2,2), and selfdual. The RR Bianchi identities are satisfied if and only if the form is closed (which implies both $W_2=0$ and co-closure, two necessary conditions). There are two such possible four-forms, as given by \eqref{eq:(2,2)NL1} and \eqref{eq:(2,2)L1}. The closure  constraint is given by \eqref{eq:(2,2)NL2}, \eqref{eq:(2,2)L2} respectively for each form. Considering that $W_4 \sim W_5$, $W_1 = W_3 = 0$, we find that we can only find non-CY solutions by ensuring $W_4 \neq 0$. Thus, our vacua will be complex but not symplectic.
\\
\\
We set $\b = r$ to satisfy $W_2 = 0$ and $r = \tanh\left(\frac{\tau}{2}\right)$ to satisfy \eqref{eq:geo}. Using this, the closure condition can be solved. Let us first examine the form which was non-$L^2$ on Stenzel space, whose closure condition is given by  \eqref{eq:(2,2)NL2}. Using our parametrization \eqref{eq:para} and the complex structure, we can solve this equation to find
\al{
f_{NL} =& m \exp(- \int \frac{a}{2b} ) \exp(- \int \frac{2b'}{b}) \exp(- \int \frac{(bc)'}{bc} )  \nn\\
=& m \frac{\a^3} {c^4 \tanh^3\left(\frac{\tau}{2} \right)\sinh\left(\frac{\tau}{2} \right) } \;,
}
with $c$ determined by \eqref{eq:susysol}. At this point $\a$ is still a free function, but we insist that $a \neq 0, \infty$ anywhere. Hence we see that regardless of choice of $\a$, supersymmetry, the Bianchi identities and the bolt at the origin do not allow for regular solutions of this flux, and the IR problems that arose for this form in Stenzel space are still there.

Let us now examine the other four-form, which behaved well on Stenzel space. The closure condition is given by \eqref{eq:(2,2)L2}, which is the same as \eqref{eq:(2,2)NL2} but with $a \leftrightarrow b$. Thus, the solution is given by
\al{
f_L =& m \exp(- \int^\tau dt \frac{b}{2a} ) \exp(- \int^\tau dt \frac{2a'}{a}) \exp(- \int^\tau dt \frac{(ac)'}{ac} )  \nn\\
=& m \frac{\a^3} {c^4 \cosh\left(\frac{\tau}{2} \right)} \;.
}
Using this expression, the warp factor can be calculated, which satisfies
\al{
\nabla^2 \hcal = - 12 g_s^2 f_L^2\;.
}
The solution is found to be
\al{
\hcal =  \frac32 g_s^2 m^2 \int_\tau^\infty dt \;  \left(\frac{\a(t)}{c(t)}\right)^6 \tanh\left(\frac{t}{2}\right) \;,
}
where an integration constant has been fixed to ensure regularity of $\hcal$ at $\tau = 0$.
Our sole free function $\a$ now needs to be chosen such that the following are satisfied:
\begin{enumerate}
\item $\a$ is nowhere vanishing nor blows up anywhere to avoid singularities. As a consequence, we immediately find that $f_L$ is regular.
\item $\hcal > 0$ holds everywhere in order to avoid the metric changing sign.
\end{enumerate}
This is a rather small list of demands, and is easily satisfied, and we will give examples below. Before doing so, however, we would like to make some remarks. The asymptotics of $\a$ govern the squashing of the metric: the $S^4$ bolt is unsquashed for $\a_{IR}=1$ and the asymptotics are precisely conical for $\a_{UV} = \sqrt{3/2}$. One might think that it is possible to use this method to also construct asymptotically $\text{AdS}_3$ rather than asymptotically conformal $\text{AdS}_3$ metrics. This is not the case. The reason is that this requires $c_{UV} \sim 1$ rather than  $c_{UV} \sim \exp(k \tau)$. Examining \eqref{eq:susysol} then leads to the conclusion that $\a_{UV} \sim \sqrt{2}$, and hence the warp factor blows up (or becomes negative, depending on the choice of integration boundaries).

Finally, before moving on to the examples, let us also comment on the possibility of a homogeneous solution. The homogeneous solutions obeys
\al{
\hcal \sim \int^\tau \frac{dt}{a^3 b^3 } \;.
}
Imposing \eqref{eq:geo} without any further restrictions due to susy or fluxes, we immediately find that $\hcal_{IR} =\int \frac{dt}{t^3}$ and hence the non-constant homogeneous warp factor retains its IR divergence regardless of choice of $abc$ $SU(4)$-structure.

\textbf{Example 1:}\\
Set
\al{
\a =  \sqrt{1 + \frac12  \tanh^2\left(\frac{\tau}{2} \right)}
}
such that
\eq{\spl{\label{eq:ex1}
a =& \l \frac{\cosh^{3/4}\left(\frac{\tau}{2} \right) }{\sqrt{1 + \frac12  \tanh^2\left(\frac{\tau}{2} \right)}}\\
b =& \l \tanh\left(\frac{\tau}{2} \right) \frac{\cosh^{3/4}\left(\frac{\tau}{2} \right) }{\sqrt{1 + \frac12  \tanh^2\left(\frac{\tau}{2} \right)}}\\
c =& \l \cosh^{3/4}\left(\frac{\tau}{2} \right)
}}
and
\al{
f_L = \frac{m \left(1 + \frac12  \tanh^2\left(\frac{\tau}{2} \right)\right)^{3/2}}{\l^4 \cosh^4\left(\frac{\tau}{2} \right)}\;.
}
Thus, the warp factor is given by
\al{
\hcal &=   \frac{g_s^2 m^2  \left(55802  + 93933 \cosh \tau + 44982 \cosh 2\tau + 13923 \cosh 3\tau\right)}{198016 \l^6\cosh^{21/2} \left(\frac{\tau}{2}\right) } \;.
}
The asymptotics are given by
\al{ \def\arraystretch{1.8}
\hcal \rightarrow \left\{ \begin{array}{cc}
\frac{1630}{1547} \frac{g_s^2 m^2}{  \l^6} - \frac{3 g_s^2 m^2}{8 \l^6}\tau^2 & \phantom{aaaaa} \tau \rightarrow 0 \\
36 \sqrt{2}\frac{ g_s^2 m^2}{\l^6}e^{-\frac94 \tau}  & \phantom{aaaaaaa} \tau \rightarrow \infty \;.
\end{array}\right.
}
We thus find an asymptotically conformal AdS metric, as the scaling is analogous to all UV-finite cases discussed in section \ref{fluxstenzel}. Specifically, we find that the (string frame) metric asymptotes to
\eq{\spl{
ds^2_{UV} &= \L^2 e^{2 \rho} [ds^2(\text{AdS}_3) + ds^2(V_{5,2})] \\
ds^2(\text{AdS}_3) &=   \d\rho^2 + e^{4 \rho} ds^2(\mathbbm{R}^{1,1}) \;,
}}
where we introduced a constant $\L$, rescaled $ ds^2(\mathbbm{R}^{1,1})$,  and $\rho \equiv \frac38 \tau$.
\\
\\
\textbf{Example 2:}\\
By changing the asymptotics of $\a$, we find solutions which asymptote to conformal AdS with a squashed $V_{5,2}$ as internal space. We now give such an example.
Set
\al{
\a = \sqrt{1 + \frac13  \tanh^2\left(\frac{\tau}{2} \right)}
}
such that
\eq{\spl{ \label{eq:ex2}
a &= \l \frac{\cosh\left(\frac{\tau}{2}\right)}{\sqrt{1+ \frac13 \tanh(\frac{\tau}{2})^2 }}\\
b &= \l   \tanh\left(\frac{\tau}{2}\right) \frac{\cosh\left(\frac{\tau}{2}\right)}{\sqrt{1+ \frac13 \tanh(\frac{\tau}{2})^2 }}\\
c &= \l \cosh\left(\frac{\tau}{2}\right) \;.
}}
This leads to a flux defined by
\al{
f_L &=  \frac{m \left(1 + \frac13  \tanh^2\left(\frac{\tau}{2}\right)\right)^{3/2}}{\l^4 \cosh^5\left(\frac{\tau}{2} \right)}\;.
}
The warp factor is given by
\al{
\hcal =  \frac{g_s^2 m^2 \left(96 + 156 \cosh \tau + 75 \cosh 2 \tau + 20 \cosh 3\tau \right) }{540 \l^6 \cosh^{12}\left(\frac{\tau}{2}\right)}
}
with asymptotics
\al{\def\arraystretch{1.8}
\hcal \rightarrow \left\{ \begin{array}{cc}
 \frac{347 g_s^2 m^2}{540 \l^6} - \frac{3 g_s^2 m^2}{8 \l^6} \tau^2 &\phantom{aaaaa} \tau \rightarrow 0\\
 \frac{2048}{27} \frac{m^2 g_s^2}{\l^6}e^{- 3  \tau} & \phantom{aaaaaa} \tau \rightarrow \infty
 \end{array}\right.
}
The metric then asymptotes to the following:
\eq{
ds^2_{UV} = \L^2 e^{\frac32\sqrt{2} \rho} [ds^2(\text{AdS}_3) + \widetilde{ds^2(V_{5,2})}]
}
where we introduced a (different) constant $\L$, the radial direction is now related to $\tau$ as $\rho = \frac{1}{\sqrt{8}} \tau$, we again rescaled $ds^2(\mathbbm{R}^{1,1})$, but the internal and external metrics are now given by
\eq{\spl{
ds^2(\text{AdS}_3) &= \d \rho^2 + e^{3 \sqrt{2} \rho} ds^2(\mathbbm{R}^{1,1})\\
\widetilde{ds^2(V_{5,2})} &= \frac12 \n^2 + \frac{3}{8} \left( \s_j^2 + \tilde{\s}_j^2\right) \;.
}}
Note that the squashed $V_{5,2}$ metric cannot simply be rescaled to the regular $V_{5,2}$ metric, due to the finite range of the (angular) coordinates. Thus, this solution is inequivalent to the first example.
\\
\\
To summarize, these solutions satisfy the following properties:
\begin{itemize}
\item The vacua are $\ncal = (1,1)$ type IIA solutions.  Due to the torsion classes, no supersymmetry enhancement occurs, and the vacua do not uplift to  $\ncal = 1$ on $\mathbbm{R}^{1,2} \times \scal$ M-theory vacua, as is evident from the fact that \eqref{eq:mtorsion} is not satisfied.
\item The RR flux is given by a closed primitive (2,2)-form.
\item $W_1 = W_2=0$ hence the almost complex structure defined by the $SU(4)$-structure is integrable.
\item $W_4 \neq 0$ hence the almost symplectic structure is not integrable. In particular, the solution is not CY.
\item $\scal$ has an $S^4$ bolt at the origin.
\item The external metric is conformally AdS for $\tau \rightarrow \infty$.
\item The RR Bianchi identity is satisfied. The NSNS Bianchi identity is not, indicating the presence of NS5-branes which act as sources.
\end{itemize}
Thus, we have shown how to break the CY structure of Stenzel space to find a vacuum that consists of a complex space and a $(2,2)$-form, at the price of introducing sources that violate the Bianchi identity for the NSNS flux.

\subsection{Source-action for vacua on complex manifolds with $f^{(2,2)}$ }\label{sources}
As mentioned, the supersymmetric vacua of the previous section violate the Bianchi identity for $H$. The source of this violation can be interpreted as a distribution of NS5-branes, as $\d H$ is not localized on $\scal$. In \cite{koerbt}, it was shown that for any violation of the RR Bianchi, an appropriate source-action can be found such that supersymmetry combined with the specific violation of RR Bianchi identity gives the source-modified equations of motions. When the NSNS Bianchi is violated instead, the situation is more complicated. This is due to the fact that the democratic supergravity action is not used to compute equations of motions for the RR charges; they are already incorporated in the Bianchi identities. Thus, the contribution to the equations of motion from the bulk supergravity action does not change, and one can look purely for a source-action to compensate the violated Bianchi identiy. This is no longer the case when $\d H \neq 0$, as will be explained below. We construct a source-action that ensures integrability is satisfied, but disregard the fact that the contribution from the bulk action should be modified.
\\
\\
Let the action be given by $S = S_{\text{bulk}} + S_{NS5}$.
The equations of motion in our conventions are then given by\footnote{See \cite{mmlt}, appendix A.}
\eq{\spl{\label{eq:eom}
D &=  2\k^2 \frac{e^{2\phi}} {\sqrt{g_{10}}} \frac{\delta S_{NS5} }{\delta \phi}\\
E_{MN} &= \k^2 \frac{e^{2\phi}} {\sqrt{g_{10}}}\left(\frac12 \frac{\delta S_{NS5} }{\delta \phi}g_{MN} - 2 \frac{\delta S_{NS5} }{\delta g^{MN}}\right) \\
\delta H_{MN} &= 2 \k^2 \frac{\delta S_{NS5}}{\delta B^{MN}}\;,
}}
with
\eq{\spl{
D &\equiv 2 R - \frac{1}{3!}H_{MNP} H^{MNP} + 8 \left(\nabla^2 \phi - (\p \phi)^2 \right) \\
E_{MN} &\equiv R_{MN} + 2 \nabla_M \nabla_N \phi - \frac{1}{2} H_{M}\cdot H_N - \frac{1}{4}e^{2 \phi} \fcal_M \cdot \fcal_N \\
\delta H_{MN} &\equiv \star_{10} e^{2 \phi} \left(\d  \left(e^{-2 \phi} \star_{10} H\right) - \frac{1}{2} \left(\star_{10}\fcal \wedge \fcal \right)_8 \right) \;.
}}
The two subtleties that are not taken into account here are as follows. Firstly, the equation of motion for $B$ has been obtained by setting $H = \d B$ in $S_{\text{bulk}}$. When $\d H \equiv j \neq 0$, this cannot be the case. Secondly, without sources, the RR equations of motion $\d_H \star \sigma \fcal = 0$ follow from the Bianchi identity $\d_H \fcal = 0$ and selfduality (see \eqref{eq:selfduality}). In case $\d H \neq 0$, $\d_H^2 \neq 0$ and hence one cannot use $\d_H \fcal = 0 \Rightarrow \fcal = \d_H C$.
\\
\\
Manipulating the supersymmetry equations, one can obtain the following integrability equations \cite{mmlt}:
\al{\label{eq:integ}
0 &= \left(- E_{MN} \G^N + \frac{1}{2} \left(\delta H_{MN} \G^N + \frac{1}{3!} (\d H)_{MNPQ} \G^{NPQ} \right)\right) \epsilon_1 - \frac{1}{4} e^{\phi} \underline{\d_H \fcal} \G_M \G_{11} \epsilon_2 \nonumber \\
0 &= \left(- E_{MN} \G^N - \frac{1}{2} \left(\delta H_{MN} \G^N + \frac{1}{3!} (\d H)_{MNPQ} \G^{NPQ} \right)\right) \epsilon_2 - \frac{1}{4} e^{\phi} \underline{\sigma \d_H \fcal} \G_M \G_{11} \epsilon_1  \nonumber \\
0 &= \left(-\frac{1}{2} D + \underline{\d H} \right) \epsilon_1 + \frac12 \underline{\d_H \fcal} \epsilon_2 \nonumber \\
0 &= \left(-\frac{1}{2} D - \underline{\d H} \right) \epsilon_2 + \frac12 \underline{\sigma \d_H \fcal} \epsilon_1
}
In our case, we have $\d_H \fcal= 0$. The spinor ansatz for $\epsilon_{1,2}$  for our solution is given in \eqref{eq:iiaks}.
The violation of the NS Bianchi identity is purely internal, i.e., the only non-vanishing components of $\d H_{MNPQ}$ are given by $\d H_{mnpq}$.
A priori, one can decompose $\d H_{mnpq}$ like any other four-form (see \eqref{eq:decomp}) as
\all{
\d H_{mnpq} =& 6 j_0 J_{[mn} J_{pq]} + 6 \left( j^{(1,1)}_{[mn} + j^{(2,0)}_{[mn}  + j^{(0,2)}_{[mn} \right)J_{pq]} \\
& + \tilde{\jmath}\; \O_{mnpq} + \tilde{\jmath}^* \O^*_{mnpq} + j^{(3,1)}_{mnpq} + j^{(1,3)}_{mnpq} \;.
}
Finally, we know that the metric satisfies $g_{m \n} = 0$, and we use the gamma matrix decomposition of \cite{pta}.
Plugging all of the above into the integrability equations, we find
\eq{\spl{\label{eq:integresults}
D &= 32 \tilde{\jmath} = 32  \tilde{\jmath}^* \\
\delta H_{MN} &= 0\\
E_{\m \n} &= 0\\
E_{mn} &= 2 (\tilde{\jmath} + \tilde{\jmath}^* ) - \frac{1}{4!}\left( j^{(3,1)}_{qrsm} \O_n^{*qrs} + j^{(1,3)}_{qrsm} \O_n^{qrs} \right)\\
j_0 &= j^{(1,1)} = j^{(2,0)} = 0\;.
}}
Our task is to figure out a suitable submanifold $M_6$ for the NS5-brane to wrap such that the contribution of the action of the NS5-brane evaluated on $M_6$ as described in \eqref{eq:eom} is equivalent to what the integrability equations tell us in \eqref{eq:integresults}, i.e.,
\eq{\spl{
2\k^2 \frac{e^{2\phi}} {\sqrt{g_{10}}} \frac{\delta S_{NS5} }{\delta \phi} &= 16 (\tilde{\jmath} + \tilde{\jmath}^* ) \\
\k^2 \frac{e^{2\phi}} {\sqrt{g_{10}}}\left(\frac12 \frac{\delta S_{NS5} }{\delta \phi}g_{mn} - 2 \frac{\delta S_{NS5} }{\delta g^{mn}}\right) &=
2 (\tilde{\jmath} + \tilde{\jmath}^* ) - \frac{1}{4!}\left( j^{(3,1)}_{qrsm} \O_n^{*qrs} + j^{(1,3)}_{qrsm} \O_n^{qrs} \right)
\\
\k^2 \frac{e^{2\phi}} {\sqrt{g_{10}}}\left(\frac12 \frac{\delta S_{NS5} }{\delta \phi}g_{\m\n} - 2 \frac{\delta S_{NS5} }{\delta g^{\m\n}}\right) &= 0 \\
2 \k^2 \frac{\delta S_{NS5}}{\delta B^{MN}} &= 0 \;.
}}
The action of the NS5-brane is known \cite{bns}, but is rather intimidating. Instead, we will simply try to construct a suitable action from scratch satisfying the above. We find that such a suitable action is given by\footnote{Taking the derivative of $\O^{mnpq}$ with respect to $g^{mn}$ may seem somewhat ambiguous, since indices could be raised with either $-i J^{mn}$ or $g^{mn}$. One should either consider all indices raised with $(\Pi^+)^{mn}$, which treats $J$ and $g$ on equal footing, or as raised with vielbeins acting on the spinor bilinear $\tilde{\eta}\g^{abcd} \eta$ with flat indices. Both give the same correct factor.}
\al{
S_{NS5} = \frac{-1 }{4!4 \k^2} \int d^{10} x \sqrt{-g_{10}}\; e^{-2 \phi} dH_{mnpq} \left( \O^{mnpq} + \O^{*mnpq} \right) \;.
}
This action has no clear DBI or WZ terms. Instead, it can be written as
\all{
S \sim \int_{\mcal_{10}} j \wedge  \Psi  \;,
}
with source $j = \d H$ and we have defined a form $\Psi$ satisfying
 \eq{\spl{
 \Psi &= \widetilde{\text{vol}_2} \wedge  \left[\text{Re}(\Omega) + ...\right]\;.
 }}
The dots represent terms that drop out of the action and the volume form $\widetilde{\text{vol}_2}$ is warped. This seems analogous to the description as given in section \ref{caliDbranes}, where calibrated D-branes were discussed. Hence we would conjecture that this action describes a calibrated distribution of NS5-branes. As far as the author is aware, no analysis is known for calibrated NS5-branes. Such an analysis for NS5-branes is more complicated than for D-branes due to a more complicated action and due to the breaking of the generalized complex geometrical framework, for which $\d H=0$ is essential.

\begin{figure}[h]
\centering
\includegraphics[scale=0.5]{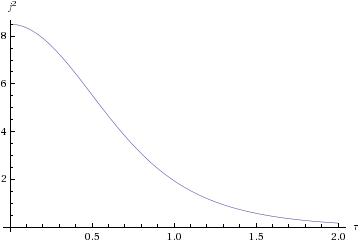}
\caption{The norm $j^2 \equiv \star (\d H \wedge \star \d H)$ as function of $\tau$ with $\l= 1$ for the first example.}
\label{fig1}
\end{figure}

\begin{figure}[h]
\centering
\includegraphics[scale=0.5]{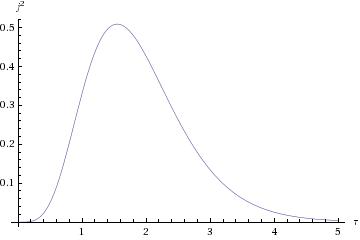}
\caption{The norm $j^2$ as function of $\tau$ with $\l= 1$ for the second example.}
\label{fig2}
\end{figure}

Let us examine the source term in more detail. Generically, it is given by the first line of \eqref{eq:h3}: the solutions we consider in section \ref{cplx22} have $W_2=0$ and thus
\al{
\d H  = \tilde{\jmath} \;\z^{0} \wedge \z^1 \wedge \z^2 \wedge \z^3   +  j_{31} \bar{\zeta}^{\bar{0}} \wedge \z^1 \wedge \z^2 \wedge \z^3 + \text{c.c.} \;,
}
with
\eq{\spl{
\tilde{\jmath} &= - \frac{1}{4 abc} \left(  3 (a^2 + b^2)w + 6 (ab)' w + 4 ab w' \right)\\
j_{31} &= - \frac{1}{4 abc} \left( - 3 (a^2 + b^2)w + 6 (ab)' w + 4 ab w' \right)\\
w &\equiv \frac12 \frac{1}{abc}\left((ab)'- \frac12 c^2 \right) \;.
}}
Thus indeed, $\d H$ is primitive with only $(4,0)$, $(3,1)$, $(1,3)$, $(0,4)$ parts and $\tilde{\jmath}$ is real, as required by \eqref{eq:integresults}.
The norm of the source is given by
\al{
j \wedge \star_8 j = 32 \left(\tilde{\jmath}^2 + j_{31}^2\right) \text{vol}_8 \;.
}
As $W_4 = 0$ if and only if $w =0$, it can be concluded that generically, $j$ will vanish for large $\tau$ due to the conical (and thus, CY) asymptotics. To be more specific, we examine the examples given in the previous section.
\\
\\
\textbf{Example 1:}\\
Imposing \eqref{eq:ex1} leads to
\al{
\star_8 \left(j \wedge \star_8 j\right) =  \frac{8863 + 6816 \cosh \tau + 1308 \cosh 2 \tau + 288 \cosh 3 \tau + 333 \cosh 4 \tau }{8 \l^4 \cosh^7 \left( \frac{\tau}{2}\right) \left(1 + 3 \cosh \tau\right)^4}
\;.
}
Thus the distribution is smooth, does not blow up, is maximal at $\tau = 0$ with norm $17 / 2 \l^4$, and falls off rapidly. The norm of the source is plotted in figure \ref{fig1} for $\l=1$.
 \pagebreak
\\
\\
\textbf{Example 2:}\\
Imposing \eqref{eq:ex2} leads to
\al{
\star\left(j \wedge \star j\right) =
\frac{8 }{9 } \frac{\tanh^4\left(\frac{\tau}{2} \right)\left(553 + 714 \cosh \tau + 278 \cosh 2 \tau + 62 \cosh 3 \tau + 13 \cosh 4\tau\right)}
{\l^4 \cosh^4\left(\frac{\tau}{2} \right)\left(1 + 2 \cosh \tau\right)^4}  \;.
}
Again, the distribution is smooth, does not blow up, and falls off rapidly. It peaks at $\tau \simeq 1.549$ with norm $\simeq 0.5090/ \l^4$. One can thus consider it to be `smeared'. The norm of the source is plotted in figure \ref{fig2} for $\l = 1$.

\section{Conclusion}
We have discussed flux compactifications on Stenzel space, constructed families of $SU(4)$-structures on $T^*S^4$, and found type IIA $\ncal= (1,1)$ compactifications on complex non-CY manifolds with primitive (2,2) RR flux which are not upliftable to M-theory vacua on $\mathbbm{R}^{1,2} \times \scal$. The latter required the introduction of NS5-brane sources. The biggest obstruction to finding more general solutions are the following: firstly, finding solutions to the RR Bianchi identities has proven to be a hurdle, independent of compactification manifold. Secondly, the fact that on Stenzel space we were unable to find a non-trivial primitive closed three-form led to a violation of the NSNS Bianchi identity for IIA vacua, and made it impossible to turn on $H$ (and consequentially, $\fcal_5$) for IIB vacua without violating all Bianchi identities. Thirdly, the family of left-invariant $SU(4)$-structures generically has $W_1 = W_3 = 0$, which meant that we were forced to consider the IIA solution branch that did not uplift to M-theory on $\mathbbm{R}^{1,2} \times \scal$, and coupled torsion classes to $H$, leading to the violation of the NSNS Bianchi identity. Finally, because it is not clear how / if it is possible to ensure geodesic completeness of the metric in another way than by using a (squashed) $S^4$ bolt, we restricted ourselves to the situation which was equivalent to Stenzel space, i.e., with conical asymptotics and an $S^4$ bolt at the origin.

All but the first of these might be alleviated by simply considering another internal manifold. The ingredients we require for the procedure applied in this paper are present on any manifold that is an asymptotically conical Calabi-Yau over a coset. The Calabi-Yau-structure grants an $SU(4)$-structure to deform, and the coset structure of the base space ensures a good control over forms by means of left-invariant forms. It would thus be interesting to repeat this procedure on other manifolds to create similar $SU(4)$-structure manifolds with non-vanishing intrinsic torsion. Similar work for heterotic vacua on non-Calabi-Yau asymptotically conical metrics using $G_2$- and $Spin(7)$-structures was done in \cite{yasui}. Non-K\"{a}hler $SU(3)$-structure deformations of the resolved conifold were found in \cite{dasgupta}.

Another possible interesting way to proceed from here is to go beyond the insistence of left-invariance of the $SU(4)$-structure. This comes down to allowing angular dependence instead of purely radial dependence of the forms, and thus becomes computationally difficult; for instance, the warp factor equation becomes a second-order PDE instead of an ODE as is the case here.
Another open question is whether or not metrics without the restriction \eqref{eq:geo} can be meaningful.

Of course, the most obvious query that arises is what the dual conformal field theory is to the $SU(4)$-deformed vacua. The Stenzel space extension has IR-singularities and should therefore be something akin to \cite{kt}. What do we find when looking at the dual to these non-CY spaces? What does the $SU(4)$-deformation translate into?
It would be interesting to have an answer to these questions.
\\
\\
{\textbf{Acknowledgements}}\\
This project started as a collaboration with C. Nun\~{e}z and D. Tsimpis, who both made contributions to the non-erroneous parts of this paper for which I am thankful. My gratitude to D. Tsimpis as well for general guidance and proofreading an early draft. I would also like to thank J.-B. Flament and S. Theisen for valuable comments and A. Tomasiello for enlightening discussion on NS5-branes.

\pagebreak

\appendix

\section{$\ncal=(2,0)$ IIB on Stenzel Space and $SU(4)$-structure deformed Stenzel Space}\label{sec:iib}
In section \ref{sec:iia} an analysis is given of $\ncal = (1,1)$ IIA supergravity on Stenzel space. We wish to give a similar analysis of $\ncal = (2,0)$ IIB supergravity on Stenzel space. However, before proceeding, a generic analysis of such IIB vacua is first required, including integrability and Bianchi identities; this is along the lines of the analysis given in \cite{pta} for IIA.

\subsection{$\ncal=(2,0)$ IIB on conformal Calabi-Yau fourfolds}\label{sec:ccy}
A general IIB $\ncal=(2,0)$ solution on CCY fourfolds is as follows. The solution presented here is obtained from \eqref{eq:iib} by imposing the following:
\eq{\spl{
\t =& \pi\\
e^{\phi} =& g_s e^{-2A}\\
f_3^{(2,1)} =& 0 \;.
}}
This leads to torsion classes
\eq{\spl{\label{eq:iibtors}
W_1 = W_2 = W_3 &= 0 \\
W_4 = \frac12 W_5 &= - 2 \p A \;,
}}
which indeed gives us a CCY metric on the internal space, related to a CY metric $g_{CY}$ by
\al{\label{eq:ccymetric}
ds^2(\mcal_8) = e^{-2 A } ds^2_{CY}(\mcal_8)\;.
}
The NSNS three-form is given by
\al{\label{eq:iibnsfluxes}
H = h^{(2,1)} + h^{(1,2)}\;,
}
in particular, $H$ is internal and primitive.
The non-vanishing RR fluxes are given by
\eq{\spl{\label{eq:iibrrfluxes}
{g_s}\fcal_3&= \mathrm{vol}_2\wedge\d e^{4A}\\
{g_s}\fcal_5&= e^{4A}\mathrm{vol}_2\wedge H- e^{2 A}\star_8 H\\
g_s \fcal_7 &= \star_{10} \s \fcal_3\;.
}}
So far, these are all consequences of supersymetry of the vacuum. Let us now consider the Bianchi identities and the integrability to a full solution to the equations of motion. The integrability theorem reviewed in \cite{pta} implies that a supersymmetric solution supplemented by the Bianchi identities and $\delta H_{01}=0$ leads to a proper vacuum. Unlike in the case of IIA in section \ref{sec:iia}, $\delta H_{01}$ is trivial in this case and leads to no additional constraints. Instead, in IIB, the constraint on the warp factor follows from the Bianchi identities.
The non-trivial Bianchi identies are given by
\eq{\spl{\label{eq:iibwarp}
\d H  &= 0 \\
\d \fcal_5 + H \wedge \fcal_3 &= 0\\
\d \fcal_7 + H \wedge \fcal_5 &= 0 \;.
}}
The second line implies
\al{ \label{eq:iibbianchis}
\d H = \d e^{2 A}\star_8 H = 0
}
while the third implies
\al{
\d \star_8 \d e^{2 A} +\frac{1}{2} H \wedge e^{2 A} \star_8 H = 0 \;.
}
In terms of the CY metric, these can be rewritten as respectively
\eq{
\d H = \d \star_{CY} H = 0
}
and
\al{\label{eq:iibwarp}
- \d \star_{CY} \d e^{-4 A} + H \wedge \star_{CY} H = 0 \;.
}

\subsection{Vacua on Stenzel Space}\label{sec:iibstenzel}
Let us specialize the solution of section \ref{sec:ccy} to the case where $\mcal_8$ is the Stenzel space $\scal$ with the appropriate CY metric and conformal factor $0$.
We have been unable to construct closed and co-closed primitive three-forms on Stenzel space so we will set $H=0$ and thus $\fcal_5=0$ as follows from the second line of (\ref{eq:iibrrfluxes}). The torsion classes all vanish for Stenzel space, hence \eqref{eq:iibtors} implies
\al{
\d A = 0\;,
}
leading to a constant warp factor. As a result, all RR fluxes vanish as well, leading to a fluxless vacuum.
In the UV, the metric asymptotes to
\al{
ds^2_{UV} = \L^2 e^{ 2 \rho} \left[ \left(e^{-2 \rho} ds^2 (\mathbbm{R}^{1,1} ) + \d\rho^2 \right) + ds^2 (V_{5,2})\right]\;,
}
with $\rho, \L$ defined in \eqref{eq:confadspara}.

\subsection{Vacua on $SU(4)$-structure deformed of Stenzel space}\label{sourcediib}
We now consider $\scal$ with a different $SU(4)$-structure. The IIB solution that we have discussed is CCY, which, together with the requirement that we have an $S^4$ bolt at the origin, leads to the conclusion that we can only consider $\scal$ as a CCY conformal to Stenzel space, as discussed in section \ref{moduli} around \eqref{eq:stenzelccy}. Specifically, we have $b, c $ fixed in terms of $a$ as
\eq{\spl{
b &=  \tanh\left(\frac{\tau}{2}\right) a \\
c &= \sqrt{\frac{2 + \cosh\tau }{3 \cosh^2\left(\frac{\tau}{2}\right)}}a \;.
}}
As $W_{4,5} \neq 0$, the warp factor is no longer constant. Thus, from \eqref{eq:iibrrfluxes}, we find that $\fcal_{3,7} \neq 0$. Therefore, the torsion constraint \eqref{eq:iibtors} and the last line of the Bianchi identities \eqref{eq:iibbianchis} impose
\eq{\spl{
- \d A &= \text{Re} W_4\\
\nabla^2 e^{-4 A} &= 0 \;.
}}
Spelled out in terms of  $a$ and $A$, these constraints are given by
\eq{\spl{
A' &=  - \frac{1}{2\tanh\left(\frac{\tau}{2}\right) a^2 } \left( (\tanh\left(\frac{\tau}{2}\right) a^2)' - \frac12 \frac{2 + \cosh\left(\tau\right) }{3 \cosh^2\left(\frac{\tau}{2}\right)} a^2 \right)\\
\p_\tau \left(\tanh^3\left(\frac{\tau}{2}\right) a^6 e^{-4 A} A'\right) &= 0 \;.
}}
The first equation can be solved explicitly to find that
\all{
a^2 = e^{-2A} \l^2 \left(2 + \cosh\tau\right)^{1/4}  \cosh\left(\frac{\tau}{2}\right) \;,
}
which is a consistency check of the fact that the metric under consideration is indeed CCY, i.e., it confirms  \eqref{eq:ccymetric} combined with \eqref{eq:stenzelabc}.
Inserting this into the second equation yields an equation for the warp factor, solved by
\al{
e^{-10 A} &=   k_1 - k_2 \int^\tau \frac{dt}{\left(x \sinh\left(\frac{t}{2}\right)\right)^3} \;,
}
with $k_{1,2}$ integration constants. Comparison with \eqref{eq:homo} leads to the conclusion that $\hcal^5$ for the IIB CCY vacuum is equivalent to the homogeneous warp factor for the IIA Stenzel space vacuum and is thus singular in the IR unless trivialized.

\section{RR-Sourced IIA Solutions}
As discussed in section \ref{rrbianchi}, we have been unable to find flux configurations with non-zero scalars or two-form on an $SU(4)$-deformed Stenzel space  that satisfy the RR Bianchi identities. We can violate the RR Bianchis at the cost of introducing more sources. These sources can be determined analogous to section \ref{sources}, but now with both RR and NSNS sources. The benefit is that, in this case, there is no other constraint than susy on the RR fluxes at all. A choice of RR fluxes then determines the warp factor. As the RR fluxes are then independent of $a,b,c$, one also has the freedom to tailor the geometry to one's wishes. We will consider this scenario here briefly to illuminate some more possibilities for the geometry.
\\
\\
\\
The IIA constraint is the torsion constraint \eqref{eq:iiasusycons}, solved by \eqref{eq:susysol}. This leaves two free functions $a,b$. In particular, these determine two interesting features to consider: the torsion and the boundary.

Let us first consider the boundary conditions we wish to impose. As before, we wish to have a (squashed) $S^4$ bolt at the origin, which means that either $(\a(0), \b(0)) = (a_0,0)$ or $(a(0),b(0)) = (0, b_0)$, $a_0, b_0 \neq 0$ and the squashing determined by the proportionality of $c(0)$ with respect to $a_0$ or $b_0$. Before, we have considered boundary conditions such that the space is asymptotically conical. Another geodesically complete option is to put another bolt at $\tau \rightarrow \infty$. We can have either similar bolts at the origin and at infinity, or different bolts, and with possibly different squashing. Our first example interpolates between these two options, with trivial squashing on both. Our second example has similar bolts, with one bolt with a fixed squashing and the other squashing a free parameter.

The second point to consider is the torsion of the $SU(4)$-structure. There are four possibilities for the torsion classes: $W_2$ and $W_4 \sim W_5$ can both be either zero or non-zero. $W_2 = 0$ has been considered when discussing the four-form solution in section \ref{cplx22}, while $W_2 = W_4 = W_5 = 0$ is the CY case, which is necessarily Stenzel space after imposing \eqref{eq:geo}. Our first example below has $W_{2,4,5} \neq 0$ whereas our second example has $W_4 = W_5 = 0$, $W_2 \neq0$: manifolds with $W_1=W_3=W_4=W_5 = 0, W_2 \neq 0$ are also referred to as `nearly Calabi-Yau'. We reiterate that $W_2 = 0$ with a bolt at the origin if and only if $ \left(\frac{a}{b}\right)^{\pm1} = \tanh\left(\frac{\tau}{2}\right)$. On the other hand, the moduli space of nearly CY spaces is more difficult to deduce. We will only consider the case where $c$ is fixed by the susy constraint \eqref{eq:iiasusycons}.  $W_4=0$ is equivalent to
\al{
(a b)' - \frac12 c^2 = 0\;,
}
which, after imposing \eqref{eq:susysol}, is equivalent to
\al{
\check{r}' + \frac32 ( \b + \frac{1}{\b} ) \check{r} - 2 =0 \;,\qquad \check{r} \equiv \frac{\a}{\b^2}
}
where we made use of the parametrization \eqref{eq:para}. Solving this equation leads to the conclusion that, for this specific $c$, the space is nearly CY if and only if
\al{\label{eq:ncy}
\a^2 = \frac{\b \exp \left(\int^\tau dt \b + \frac{1}{\b}\right) }{k + 2 \int^\tau dt \exp \left(\int^\tau dt \b + \frac{1}{\b}\right)}\;,
}
with $k$ an integration constant.

\textbf{Example 1:}
Let us consider the case where we want $W_{2,4,5} \neq 0$ with bolts at  $\tau = 0, \infty$. In this case, it will be easiest to forego the parametrization \eqref{eq:para}.
Set
\eq{\spl{
c &= \l\\
a &= \l \cos(h(\tau)) \\
b &= \l \sin(h(\tau)) \;.
}}
It can easily be verified that this satisfies \eqref{eq:iiasusycons} and that $W_{2,4,5} \neq 0$,  for any function $h(\tau)$. Furthermore, if $h(0) = 0$, \eqref{eq:geo} is satisfied, thus leading to the usual $S^4$ bolt at the origin.
Let us set
\al{
h(\tau) = k \arctan(\tau) + (1-k) N \tau e^{- \tau} \;.
}
$k \in [0,1]$ interpolates between the solution where $c_{UV} = a_{UV} = \l, b_{UV} = 0$ for $k= 0$ and  $a_{UV} = 0, c_{UV} = b_{UV} = \l$ for $k=1$. Hence we can choose which of the two possible (non-squashed) $S^4$ bolts we have at $\tau \rightarrow \infty$. Of course, to swap bolt type at $\tau = 0$, we need simply swap $ a\leftrightarrow b$.
We choose the constant $N$ to be suitably small such that $h(\tau) \in (0, \frac{\pi}{2})$  $\forall \tau \in (0, \infty)$, hence ensuring that $a,b$ do not vanish or blow up at any other point.
By choosing a suitable warp factor, the external metric can be taken to be $\text{AdS}_3$, globally rather than just asymptotically.

\textbf{Example 2:} \\
Let us now construct a nearly CY with two similar squashed $S^4$ bolts, i.e.,
\eq{\spl{
a(0) &= a_0 \;, \quad b(0) = 0 \\
\lim_{\tau \rightarrow \infty} a(\tau) &\equiv a_{UV} \;, \quad \lim_{\tau -> \infty} b(\tau) = 0 \;.
}}
Such a solution is given by
\al{
\b = \frac{k_1 \tau}{k_2 \tau^{1 + k_3 } + 1} \;, \qquad k_{1,2,3} \in (0, \infty) \;.
}
Clearly this satisfies $b(0) = \lim\limits_{\tau -> \infty} b(\tau) = 0$. Defining $\a$ as in \eqref{eq:ncy} with $k=0$, we find that
\eq{\spl{
\a(0) &= \sqrt{\frac{3}{4} + \frac{k_1}{2}}\\
\lim_{\tau \rightarrow \infty} \a &=  \sqrt{3/4}\;,
}}
thus satisfying all boundary conditions. As the squashing of the $S^4$ is non-trivial for $\a \neq 1$, the bolt at $\tau = \infty$ has fixed non-trivial squashing whereas the squashing of the $S^4$ at the origin is determined by $k_1$.

\pagebreak


\begin{thebibliography}{99}




\bibitem{malnun}
  J.~M.~Maldacena and C.~Nu\~{n}ez,
{\it Supergravity description of field theories on curved manifolds and a no go theorem},
  Int.\ J.\ Mod.\ Phys.\ A {\bf 16} (2001) 822,
  \arx{hep-th/0007018}.

\bibitem{stenzel}
M.~Stenzel,
{\it Ricci-flat metrics on the complexification of a compact rank one symmetric space},
Manuscripta Math.80 151 (1993),
\href{https://people.math.osu.edu/stenzel.3/research/publications/ricci-flat.pdf}{https://people.math.osu.edu/stenzel.3/research/publications/ricci-flat.pdf}.

\bibitem{eguchi}
T.~Eguchi and A.~J.~Hanson,
{\it Selfdual Solutions to Euclidean Gravity},
Annals Phys.\  {\bf 120} (1979) 82.


\bibitem{candelas}
P.~Candelas and X.~C.~de la Ossa,
{\it Comments on Conifolds},
Nucl.\ Phys.\ B {\bf 342} (1990) 246.


\bibitem{kw}
I.~R.~Klebanov and E.~Witten,
{\it AdS / CFT correspondence and symmetry breaking},
Nucl.\ Phys.\ B {\bf 556} (1999) 89,
\arx{hep-th/9905104}.

\bibitem{kt}
I.~R.~Klebanov and A.~A.~Tseytlin,
{\it Gravity duals of supersymmetric $SU(N) \times SU(N+M)$ gauge theories},
Nucl.\ Phys.\ B {\bf 578} (2000) 123,
\arx{hep-th/0002159}.

\bibitem{ks}
I.~R.~Klebanov and M.~J.~Strassler,
{\it Supergravity and a confining gauge theory: Duality cascades and chi SB resolution of naked singularities},
JHEP {\bf 0008} (2000) 052,
\arx{hep-th/0007191}.




\bibitem{cglp}
  M.~Cvetic, G.~W.~Gibbons, H.~Lu and C.~N.~Pope,
  {\it Ricci flat metrics, harmonic forms and brane resolutions},
  Commun.\ Math.\ Phys.\  {\bf 232} (2003) 457,
  \arx{hep-th/0012011}.




\bibitem{kp}
I.~R.~Klebanov and S.~S.~Pufu,
  {\it M-Branes and Metastable States},
  JHEP {\bf 1108} (2011) 035,
  \arx{1006.3587}.

\bibitem{bena}
I.~Bena, G.~Giecold and N.~Halmagyi,
  {\it The Backreaction of Anti-M2 Branes on a Warped Stenzel Space},
  JHEP {\bf 1104} (2011) 120,
  \arx{1011.2195}.

\bibitem{massai}
S.~Massai,
  {\it Metastable Vacua and the Backreacted Stenzel Geometry},
  JHEP {\bf 1206} (2012) 059,
  \arx{1110.2513}.

\bibitem{hashimoto}
A.~Hashimoto and P.~Ouyang,
 {\it Quantization of charges and fluxes in warped Stenzel geometry},
  JHEP {\bf 1106} (2011) 124,
  \arx{1104.3517}.

\bibitem{ms}
  D.~Martelli and J.~Sparks,
{\it AdS(4) / CFT(3) duals from M2-branes at hypersurface singularities and their deformations},
  JHEP {\bf 0912} (2009) 017,
  \arx{0909.2036}.

\bibitem{gauntlett}
J.~P.~Gauntlett, D.~Martelli and D.~Waldram,
{\it Superstrings with intrinsic torsion},
Phys.\ Rev.\ D {\bf 69} (2004) 086002,
\arx{hep-th/0302158}.

\bibitem{ptb}
D.~Prins and D.~Tsimpis,
  {\it IIA supergravity and M-theory on manifolds with SU(4) structure},
  Phys.\ Rev.\ D {\bf 89} (2014) 064030,
  \arx{1312.1692}.


\bibitem{pta}
  D.~Prins and D.~Tsimpis,
  {\it IIB supergravity on manifolds with SU(4) structure and generalized geometry},
  JHEP {\bf 1307} (2013) 180,
  \arx{1306.2543}.

\bibitem{mmlt}
  D.~Lust, F.~Marchesano, L.~Martucci and D.~Tsimpis,
{\it Generalized non-supersymmetric flux vacua},
  JHEP {\bf 0811} (2008) 021
  \arx{0807.4540}.

\bibitem{melec}
  L.~Martucci,
  {\it Electrified branes},
  JHEP {\bf 1202} (2012) 097,
  \arx{1110.0627}.

\bibitem{nieuw}
P.~van Nieuwenhuizen,
{\it General theory of coset manifolds and antisymmetric tensors applied
to Kaluza-Klein supergravity},
Supersymmetry and supergravity 84, World Scientific.

\bibitem{varela}
D.~Cassani, P.~Koerber, O.~Varela,
{\it All homogeneous $N=2$ M-theory truncations with supersymmetric $\text{AdS}_4$ vacua},
JHEP {\bf 1211} (2012) 173,
\arx{1208.1262}.

\bibitem{hitchin}
  N.~J.~Hitchin,
{\it The geometry of three-forms in six and seven dimensions},
  \arx{math/0010054}.

  \bibitem{ckkltz}
  C.~Caviezel, P.~Koerber, S.~Kors, D.~Lust, D.~Tsimpis and M.~Zagermann,
  {\it The Effective theory of type IIA AdS(4) compactifications on nilmanifolds and cosets},
  Class.\ Quant.\ Grav.\  {\bf 26} (2009) 025014,
  \arx{0806.3458}.

\bibitem{koerbt}
P.~Koerber and D.~Tsimpis,
{\it Supersymmetric sources, integrability and generalized-structure compactifications},
  JHEP {\bf 0708} (2007) 082,
  \arx{0706.1244}.




\bibitem{bns}
  I.~A.~Bandos, A.~Nurmagambetov and D.~P.~Sorokin,
{\it The Type IIA NS5-brane},
  Nucl.\ Phys.\ B {\bf 586} (2000) 315,
  \arxth{0003169}.

\bibitem{yasui}
K.~Hinoue and Y.~Yasui,
{\it Heterotic Solutions with $G_2$ and $Spin(7)$ Structures},
  \arx{1410.7700}.

\bibitem{dasgupta}
K.~Dasgupta, M.~Emelin and E.~McDonough,
{\it Non-Kahler Resolved Conifold, Localized Fluxes in M-Theory and Supersymmetry},
  \arx{1412.3123}.






\end{thebibliography}
\end{document}